%% file: main-V14-arXiv.tex
\documentclass[10pt, final, twocolumn, twoside, romanappendices]{IEEEtran}
\usepackage{amsthm,amsmath,amsfonts,amssymb}
\usepackage{graphicx}
\usepackage[table]{xcolor}
\usepackage{stmaryrd} 
\usepackage{bm}
\usepackage{float}
\usepackage{stfloats}
\usepackage{mathrsfs}
\usepackage{bbm}
\usepackage{mathtools}  
\usepackage{textgreek}  
\usepackage{cuted}
\usepackage{color}
\usepackage{array,multirow, makecell}
\usepackage{colortbl}  
\usepackage{hhline}
\usepackage{tabulary}
\usepackage{booktabs,siunitx}
\usepackage{multicol}
\usepackage{balance}
\usepackage{xstring}
\usepackage{epstopdf}
\usepackage{epsfig,psfrag}
\usepackage{verbatim}
\usepackage{pgfplots}
\usepackage{flushend}
\usepackage{cite}
\usepackage[tight,footnotesize]{subfigure}
\usepackage[]{hyperref}
\usepackage{nameref}

\usepackage{enumitem, hyperref}
\makeatletter
\def\namedlabel#1#2{\begingroup
	#2%
	\def\@currentlabel{#2}%
	\phantomsection\label{#1}\endgroup
}
\makeatother

\begingroup%
  \makeatletter%
  \providecommand\color[2][]{%
    \errmessage{(Inkscape) Color is used for the text in Inkscape, but the package 'color.sty' is not loaded}%
    \renewcommand\color[2][]{}%
  }%
  \providecommand\transparent[1]{%
    \errmessage{(Inkscape) Transparency is used (non-zero) for the text in Inkscape, but the package 'transparent.sty' is not loaded}%
    \renewcommand\transparent[1]{}%
  }%
  \newcommand*\fsize{\dimexpr\f@size pt\relax}%
  \newcommand*\lineheight[1]{\fontsize{\fsize}{#1\fsize}\selectfont}%
  \ifx\svgwidth\undefined%
    \ifx\svgscale\undefined%
      \relax%
    \else%
    \fi%
  \else%
  \fi%
  \global\let\svgwidth\undefined%
  \global\let\svgscale\undefined%
  \makeatother%

\DeclareMathAlphabet{\mathsfbr}{OT1}{cmss}{m}{n}
\SetMathAlphabet{\mathsfbr}{bold}{OT1}{cmss}{bx}{n}
\DeclareRobustCommand{\msf}[1]{%
  \ifcat\noexpand#1\relax\msfgreek{#1}\else\mathsfbr{#1}\fi
}

\DeclareRobustCommand{\mcal}[1]{%
  \ifcat\noexpand#1\relax\mathnormal{#1}\else\cal{#1}\fi
}
\DeclareRobustCommand{\BM}[1]{%
  \ifcat\noexpand#1\relax\bm{\boldUppercaseItalicGreek{#1}}\else\bm{#1}\fi
}

\newcommand{\dx}{\mathrm{d}}
\newcommand{\fg}{f_{\scriptscriptstyle \mathrm{G}}}
\newcommand{\fgm}{f_{\scriptscriptstyle \mathrm{GM}}}

\newcommand{\rv}[1]{\msf{#1}}
\newcommand{\RV}[1]{\bm{\msf{#1}}}

\newcommand{\RS}[1]{\msf{#1}}

\newcommand{\V}[1]{\bm{#1}}
\newcommand{\M}[1]{\BM{#1}}
\newcommand{\Set}[1]{{\mcal{#1}}}

\newcommand{\card}[1]{\left\lvert#1\right\rvert}

\newcommand{\intset}[2]{\llbracket #1, #2 \rrbracket }

\input{Acronym-Def.tex}

\DeclareFontEncoding{LGR}{}{} 
\DeclareSymbolFont{sfgreek}{LGR}{cmss}{m}{n}
\SetSymbolFont{sfgreek}{bold}{LGR}{cmss}{bx}{n}
\DeclareMathSymbol{\salpha}{\mathord}{sfgreek}{`a}
\DeclareMathSymbol{\sbeta}{\mathord}{sfgreek}{`b}
\DeclareMathSymbol{\sgamma}{\mathord}{sfgreek}{`g}
\DeclareMathSymbol{\sdelta}{\mathord}{sfgreek}{`d}
\DeclareMathSymbol{\sepsilon}{\mathord}{sfgreek}{`e}
\DeclareMathSymbol{\szeta}{\mathord}{sfgreek}{`z}
\DeclareMathSymbol{\seta}{\mathord}{sfgreek}{`h}
\DeclareMathSymbol{\stheta}{\mathord}{sfgreek}{`j}
\DeclareMathSymbol{\siota}{\mathord}{sfgreek}{`i}
\DeclareMathSymbol{\skappa}{\mathord}{sfgreek}{`k}
\DeclareMathSymbol{\slambda}{\mathord}{sfgreek}{`l}
\DeclareMathSymbol{\smu}{\mathord}{sfgreek}{`m}
\DeclareMathSymbol{\snu}{\mathord}{sfgreek}{`n}
\DeclareMathSymbol{\sxi}{\mathord}{sfgreek}{`x}
\DeclareMathSymbol{\somicron}{\mathord}{sfgreek}{`o}
\DeclareMathSymbol{\spi}{\mathord}{sfgreek}{`p}
\DeclareMathSymbol{\srho}{\mathord}{sfgreek}{`r}
\DeclareMathSymbol{\ssigma}{\mathord}{sfgreek}{`s}
\DeclareMathSymbol{\stau}{\mathord}{sfgreek}{`t}
\DeclareMathSymbol{\supsilon}{\mathord}{sfgreek}{`u}
\DeclareMathSymbol{\sphi}{\mathord}{sfgreek}{`f}
\DeclareMathSymbol{\schi}{\mathord}{sfgreek}{`q}
\DeclareMathSymbol{\spsi}{\mathord}{sfgreek}{`y}
\DeclareMathSymbol{\somega}{\mathord}{sfgreek}{`w}

\DeclareMathSymbol{\svarsigma}{\mathord}{sfgreek}{`c}

\DeclareMathSymbol{\sGamma}{\mathalpha}{sfgreek}{`G}
\DeclareMathSymbol{\sDelta}{\mathalpha}{sfgreek}{`D}
\DeclareMathSymbol{\sTheta}{\mathalpha}{sfgreek}{`J}
\DeclareMathSymbol{\sLambda}{\mathalpha}{sfgreek}{`L}
\DeclareMathSymbol{\sXi}{\mathalpha}{sfgreek}{`X}
\DeclareMathSymbol{\sPi}{\mathalpha}{sfgreek}{`P}
\DeclareMathSymbol{\sSigma}{\mathalpha}{sfgreek}{`S}
\DeclareMathSymbol{\sUpsilon}{\mathalpha}{sfgreek}{`U}
\DeclareMathSymbol{\sPhi}{\mathalpha}{sfgreek}{`F}
\DeclareMathSymbol{\sPsi}{\mathalpha}{sfgreek}{`Y}
\DeclareMathSymbol{\sOmega}{\mathalpha}{sfgreek}{`W}

\def\indicatorsymbol{\text{\usefont{U}{bbold}{m}{n}{1}}}
\newcommand{\indicator}[1]{\indicatorsymbol_{#1}}

\newcommand{\dtv}{d_{\scriptscriptstyle \mathrm{TV}}}
\newcommand{\datv}{d_{\scriptscriptstyle \mathrm{ATV}}}

\newcommand{\no}{n_{\mathrm o}}

\newcommand{\Puk}{P_{{\mathrm u}, k}}
\newcommand{\Quk}{Q^{}_{{\mathrm u},k}}
\newcommand{\Vuk}[1]{V^{#1}_{{\mathrm u},k}}

\newcommand{\tPuk}{\tilde{P}_{{\mathrm u}, k}}

\newcommand{\tQuk}{\tilde{Q}_{{\mathrm u},k}}

\newcommand{\tVuk}[1]{\tilde{V}^{#1}_{{\mathrm u},k}}

\newcommand{\Au}{\M{A}_{{\mathrm u}, k}}

\newcommand{\h}{\V{h}}
\newcommand{\hv}{\V{h}_{\mathrm{r}}}

\newcommand{\fuk}[1]{f^{#1}_{\mathrm{u}, k}}
\newcommand{\fu}[1]{f^{#1}_{\mathrm{u}}}

\newcommand{\ruk}[1]{r^{#1}_{\mathrm{u}, k}}
\newcommand{\ru}[1]{r^{#1}_{\mathrm{u}}}

\newcommand{\truk}[1]{\tilde{r}^{#1}_{\mathrm{u}, k}}
\newcommand{\tru}[1]{\tilde{r}^{#1}_{\mathrm{u}}}

\newcommand{\Pdk}[1]{P^{#1}_{\rmv \mathrm{d}, k}}
\newcommand{\Pd}[1]{P^{#1}_{\rmv \mathrm{d}}}

\newcommand{\Mabt}[1]{{\Upsilon}^{(#1)}}
\newcommand{\Mbat}[1]{\Psi^{(#1)}}

\renewcommand{\card}[1]{ \vert #1 \vert }   

\newcommand{\ist}{\hspace*{.3mm}}
\newcommand{\rmv}{\hspace*{-.3mm}}

\newcommand{\nn}{\nonumber}

\makeatother

\newcommand{\ffak}{f_{\mathrm{c},k}}
\newcommand{\lfak}{\lambda_{\mathrm{c},k}}
\newcommand{\fa}{{\mathrm{c}}}
\newcommand{\ffa}{f_{\mathrm{c}}}
\newcommand{\lfa}{\lambda_{\mathrm{c}}}

\newcommand{\ie}{i.e.}
\newcommand{\eg}{e.g.}

\DeclareMathAlphabet{\mathpzc}{OT1}{pzc}{m}{it}

\makeatletter
\def\munderbar#1{\underline{\sbox\tw@{$#1$}\dp\tw@\z@\box\tw@}}
\makeatother

\newtheorem{theorem}{Theorem}[section]

\newtheorem{proposition}[theorem]{Proposition}

\begin{document}

\title{Gaussian Belief Propagation for Tracking \\With Unresolved Measurements\vspace{3mm}}

\author{Augustin~A.~Saucan,~\IEEEmembership{Member,~IEEE}, Florian~Meyer,~\IEEEmembership{Member,~IEEE}, Peter~Willett,~\IEEEmembership{Fellow,~IEEE}
	\vspace{-2.5ex}	
	\thanks{
		
    A.~A.~Saucan is with the Institute of Telecommunications, TU Wien, Vienna 1040, Austria (email: \texttt{augustin.saucan@tuwien.ac.at}). F.~Meyer is with the Scripps Institution of Oceanography and the Department of Electrical and Computer Engineering, University of California San Diego, San Diego, CA (e-mail: \texttt{flmeyer@ucsd.edu}). P. Willett is with the Department of Electrical and Computer Engineering, University of Connecticut, Storrs, CT, USA (email: \texttt{peter.willett@uconn.edu}). This work has been submitted to the IEEE for possible publication. Copyright may be transferred without notice, after which this version may no longer be accessible.}
}


\maketitle

\begin{abstract}
Unresolved measurements occur in many inference problems where two or more hidden processes may, at times, jointly generate a single measurement. For instance, such phenomena are encountered in multiobject tracking owing to the limited resolution capabilities of practical sensors; or in camera-aided autonomous driving due to shadowing or occlusions. Substantial performance degradation, such as track losses, are incurred when unresolved measurements are not accounted for.

In this paper, we address multiobject tracking under a generalized unresolved measurement model, where any subset of objects may generate a single unresolved measurement according to a probabilistic model. Our innovation lies both in modeling and algorithm-design directions. First, we develop a probability distribution for object partitions based on a model of pairwise coupling of objects and subsequently a probability distribution for object-to-measurement association variables. This generic model incorporates sensor resolution capabilities, sensor detection and noise characteristics for object groups. Second, a generic \ac{lbp} method as well as a specialized~\ac{glbp} algorithm are proposed that perform object state inference under the aforementioned model. In contrast to direct marginalization methods, which involve a computational complexity of $O(m^n)$, for $m $ measurements and $n$ objects, the proposed~\ac{glbp} algorithm achieves a computational complexity on the order of $O(m n 2^{n})$. Numerical results demonstrate the effectiveness of our proposed \ac{glbp}, with estimation performance that closely matches that of exact marginalization for only a fraction of the computational resources. 
\end{abstract}


%

\section{Introduction}

Multiobject tracking~\cite{BarLi:B95,SauChoSinCai:16,MeyBraWilHla:J17,BeaVoVo:15,SauCoaRab:17,MeyKroWilLauHlaBraWin:J18,VoVoPhu:14,SauWin:J20} is the problem of estimating the states of multiple hidden processes from available sensor measurements.
Multiobject tracking in densely cluttered environments is of considerable interest across a variety of engineering domains, including autonomous navigation, applied ocean sciences, air traffic control, and biomedical applications. In many such applications, sensors supply a set of point measurements for which the added combinatorial optimization problem of data association, i.e., unknown associations between measurements and objects, requires addressing before state inference can be carried out. 

In traditional data-association problems, a one-to-one correspondence is assumed between detected objects and measurements. However, the much harder problem of many-to-one associations arises in practical situations whenever objects occlude/shadow one another or when objects are unresolved vis-\`{a}-vis the sensor resolution capabilities. An unresolved group of objects may generate a single measurement and when not explicitly modeled, such situations can lead to inaccurate tracking results and frequent track losses. 

For monopulse radars, a maximum likelihood estimator coupled with the minimum description length criterion is employed in \cite{XinWilBSh:J05} in order to estimate the number and the location information of multiple unresolved objects. Also for monopulse radars, a particle filter approach coupled with a likelihood-ratio test is developed in \cite{IsaWilBSh:J08} in order to detect and estimate object spawning phenomena.

Several early approaches for multiobject tracking with potentially unresolved measurements address tracking of only two objects. In \cite[Ch.~9.4]{BarLi:B95} two crossing objects are tracked using imaging sensors. Image processing tools are used to detect potential unresolved situations and switch between independent and coupled tracking. Fixed-grid resolution models were employed in~\cite{ChaSha:C83,ChaSha:J84} and \cite[Ch.~6.4]{BarLi:B95} in order to detect when two objects become unresolved. Based on this model, Gaussian approximations are derived for the marginal state probability density functions (pdfs) of the two objects. The resulting method, called \ac{jpdam} in~\cite{ChaSha:C83,ChaSha:J84}, was subsequently refined in~\cite{ChaSha:J86} through a simplified approximation of the merged-measurement pdf.

Departing from the fixed-grid resolution assumption in~\cite{ChaSha:C83,ChaSha:J84,ChaSha:J86}, the authors of~\cite{KocKeu:J97} introduce a sensor resolution model in which the probability that two objects are unresolved is explicitly characterized. Tracking of two potentially unresolved objects is then addressed using a generalized multiple hypothesis tracking formulation that incorporates unresolved-object hypotheses. This resolution model is further employed in~\cite{JeoTug:J08}, where an interacting multiple model \ac{jpdam} approach is proposed to handle both maneuvering objects and unresolved measurements. Using tools from analytic combinatorics, in~\cite{AngStrEfe:J21} a joint likelihood function is derived so that it accounts for all feasible measurement-to-object association hypotheses. Based on this formulation, a \ac{jpdaf} is proposed for a two-object tracking scenario. Still based on the model of~\cite{KocKeu:J97} and on~\ac{lbp}~\cite{MeyKroWilLauHlaBraWin:J18,WilLau:J14,MeyBraWilHla:J17,MeyWil:J21}, a method that tracks multiple objects but where the merged measurement phenomenon is limited to groups of only two objects is proposed in~\cite{SauMey:C23}.  

Resolution graphs are introduced in~\cite{SveUlmHam:12} as a model for arbitrary-sized groups of objects that generate an unresolved measurement. The nodes of a resolution graph are the objects while an edge between two nodes indicates that the two objects are unresolved. Furthermore, a probabilistic model is implicitly assumed for resolution graphs since each edge is considered a Bernoulli random variable with probability of success given by the same pairwise coupling model of~\cite{KocKeu:J97}. The resulting framework can be interpreted as a generalization of the two-object resolution model given in~\cite{KocKeu:J97} to a fixed but arbitrary number of objects. Subsequently in~\cite{SveUlmHam:12}, the joint posterior pdf of objects is obtained by directly marginalizing over all valid resolution graphs and association hypotheses. In order to reduce computational complexity, a sensor-dependent strategy is employed to eliminate certain resolution graphs.

A multiobject tracking approach based on labeled random finite sets is proposed in~\cite{BeaVoVo:15}, where a \ac{glmb} density is employed in conjunction with a merged-measurement model that does not take into account any specific sensor resolution capabilities.  In a similar manner to~\cite{SveUlmHam:12}, the~\ac{glmb} filter ideally evaluates all association hypotheses between objects and measurements and, unlike~\cite{SveUlmHam:12}, constructs a tree of such hypotheses through time. In practice, however, the exponential growth of this tree needs to be mitigated by pruning techniques. An alternative approach is the merged measurement labeled multi-Bernoulli filter proposed in~\cite{SauWin:C20}; however, this still requires the construction and evaluation of a large number of association hypotheses during the update step, followed by explicit marginalization for each object across these hypotheses. Also in \cite{LiGaoZhaoWei:J25}, a multiobject tracking filter is proposed for merged measurements without any sensor resolution model.

In this work, we consider the problem of tracking a fixed but arbitrary number of objects under the assumption that any subset of objects may become unresolved. The novelty of our approach lies both in the modeling and algorithm-design directions. First, we develop a probability distribution of object partitions based on the pairwise coupling model of~\cite{KocKeu:J97} and subsequently a prior probability distribution for association variables. This generic model is flexible enough to allow the incorporation of sensor resolution capabilities, a detection process, and sensor noise characteristics for object groups; while being capable of modeling scenarios where several arbitrary-sized groups of objects generate unresolved measurements. Second, a generic \ac{lbp} method is proposed that performs object-state inference under the aforementioned model. Furthermore, this generic method is specialized to a~\ac{glbp} algorithm that employs Gaussian distributions for object posteriors and which achieves a computational complexity that scales on the order of $O(m n 2^{n})$, for $m $ measurements and $n$ objects. This is a substantial improvement over a method that relies on exact marginalization with a complexity of $O(m^n)$.

The remainder of the paper is organized as follows. Notation and the system model are introduced in Sec.~\ref{sec:model}. The proposed distribution for object partitions is presented in Sec.~\ref{sec_unresolved_model}. In Sec.~\ref{sec:posterior}, the data-association model with unresolved measurements is introduced, and together with the distribution for object partitions is used to develop the multiobject posterior distribution. A generic \ac{lbp} algorithm for multiobject tracking is derived in Sec.~\ref{sec_prop_alg1}, while a \ac{glbp} filter is proposed in Sec.~\ref{sec_prop_alg2}. Numerical results are provided in Sec.~\ref{sec:results}.

\section{Notation and System Model} \label{sec:model}
Throughout this work, the following notations are employed. Random variables are displayed in sans serif, upright fonts; their realizations in serif, italic fonts. Vectors and matrices are denoted by bold lowercase and uppercase letters, respectively. Random sets and their realizations are denoted by upright sans serif and calligraphic font, respectively. For example, a random vector and its realization are denoted by $\RV{x}$ and $\V{x}$ while a random set and its realization are denoted by $\RS{X}$ and $\Set{X}$, respectively. The set cardinality is denoted by $\card{\RS{X}}$. The set of integers $\{ a, a+1, \cdots, b\}$, for any $a<b$, is denoted as $\intset{a}{b}$.  A set indicator function is introduced as $\indicator{\Set{X}}(\Set{Y}) = 1$ if $\Set{Y} \subseteq \Set{X}$ and $\indicator{\Set{X}}(\Set{Y}) = 0$ otherwise. A set equality operator is defined as $\delta_{\Set{X}}(\Set{Y}) = 1$ if $\Set{X} = \Set{Y}$ and $\delta_{\Set{X}}(\Set{Y}) = 0$ otherwise. The product of a function $h(\cdot)$ over all the elements of a set $\Set{X}$ is denoted as $\prod_{x\in \Set{X}} h(x)$. We make the convention of $\prod_{x\in \Set{X}} h(x) = 1$ when $\Set{X}=\emptyset$. Equality up to a normalization factor is denoted as $\propto$.

An undirected graph $G$ is defined by the set of vertices/nodes $\Set{V} \subset \mathbb{N}$ and the set of edges $\Set{E}  \subset \Set{V}  \times \Set{V}$. The set of vertices of a graph $G$ is denoted via $\Set{V}_{G}$ and the set of edges via $\Set{E}_{G}$. For a graph $G$, we say two of its vertices $a,b \in \Set{V}_{G}$ are connected when $G$ contains a path between the two vertices, which we denote by $a \leftrightsquigarrow b  $. Trivially $a \leftrightsquigarrow a  $ for any $a \in \Set{V}_{G}$. For a given set of vertices $\Set{A} \subset \mathbb{N}$, the set of all graphs spanning the vertices in $\Set{A}$ is denoted via $\mathfrak{G}(\Set{A})$. The set of all partitions of a set $\Set{A}$ is denoted via $\mathfrak{B}(\Set{A})$.

 A component of an undirected graph $G$ is defined as a connected subgraph of $G$ that is not part of any larger connected subgraph \cite{Die:B17}. Let $\Set{C}(G)$ be the union of component vertex sets of $G$. More specifically, $\Set{C}(G) \triangleq \{ C(v) \colon v\in \Set{V}_G\}$ and $C(v) \triangleq \{u\in \Set{V}_G \colon u \leftrightsquigarrow v\}$. Note that the elements of $\Set{C}(G) $ effectively partition the graph vertices, i.e., $\Set{C}(G)  = \{ C_i \colon \cup_i {C_i} = \Set{V}_G,\, {C_i} \cap {C_j} = \emptyset \text{ for } i\neq j \}$.

 
 A Gaussian probability density function with variable $\V{x}\in \mathbb{R}^d$, mean vector $\V{m}\in \mathbb{R}^d$, and covariance matrix $\M{P}\in \mathbb{R}^{d\times d}$ is denoted as $\fg(\V{x}; \V{m}, \M{P})$. For some natural $n\geq 1$, let $\munderbar{\V{w}} = (w_1, w_2,\cdots, w_n)$ be a tuple of non-negative weights that sum to one, $\munderbar{\V{m}} = (\V{m}_1, \V{m}_2, \cdots, \V{m}_n)$ be a tuple of vectors from $\mathbb{R}^d$, and $\munderbar{\M{P}}=(\M{P}_1, \M{P}_2, \cdots, \M{P}_n)$ a tuple of covariance matrices from $ \mathbb{R}^{d\times d}$, then we denote a Gaussian mixture via $\fgm(\V{x}; \munderbar{\V{w}}, \munderbar{\V{m}}, \munderbar{\V{P}}) \triangleq \sum_{i=1}^{n} w_i \fg(\V{x}; \V{m}_i , \M{P_i})$. A projection of a Gaussian mixture onto the space of single Gaussian probability densities that preserves the first and second order moments of the original mixture is defined as 
\begin{equation}
(\Pi \fgm) (\V{x}) = \fg (\V{x}; \V{m}, \M{P})
\label{eq_gm_proj}
\end{equation}
where $\V{m} = \sum_{i=1}^n {w}_i \V{m}_i $ and $\M{P}  =\sum_{i=1}^n {w}_i \M{P}_i + \sum_{i=1}^{n} {w}_i (\V{m}_i - \V{m}) (\V{m}_i - \V{m})^\top$.

\subsection{System Model}

 A number $\no$ of mobile objects is considered. The trajectory of each object $i\in \Set{O} \triangleq \intset{1}{n_{\mathrm{o}}}$ is described by a stochastic process $\RV{x}^{(i)}_k \in \mathbb{X}$ indexed by the discrete time $k$ and taking values in the state space $\mathbb{X} \subseteq \mathbb{R}^{d_{\rm x}}$. The multiobject stochastic process is given by $\munderbar{\RV{x}}_k   = \big[ (\RV{x}^{(1)}_k)^\top, (\RV{x}^{(2)}_k)^\top, \cdots, (\RV{x}^{(\no)}_k)^\top \big]^\top$\rmv\rmv\rmv\rmv. At time $k$, a sensor produces a random number $\rv{m}_k \in \mathbb{N}$ of measurements that is collected in a single vector $\munderbar{\RV{z}}_k = \big[ (\RV{z}^{(1)}_k)^\top, (\RV{z}^{(2)}_k)^\top, \cdots, (\RV{z}^{(\rv{m}_k)}_k)^\top \big]^\top$\rmv\rmv\rmv, where each individual measurement is $\RV{z}_k^{(j)} \in \mathbb{Z}$, with $ \mathbb{Z} \subseteq \mathbb{R}^{d_{\rm z}}$. The collection of all measurements up to and including time $k$ is denoted via $\munderbar{\RV{z}}_{1:k} \triangleq (\munderbar{\RV{z}}_1, \munderbar{\RV{z}}_2, \cdots, \munderbar{\RV{z}}_k)$. Furthermore, for any subset of indices $\Set{I} \triangleq \{i_1, i_2, \cdots, i_m\} \subset \Set{O}$, the set of corresponding object state vectors $\big( \RV{x}^{(i_1)}_k, \RV{x}^{(i_2)}_k, \cdots, \RV{x}^{(i_m)}_k \big)$ is denoted via $\munderbar{\RV{x}}_k^{\Set{I}}$. 
 
 Considering the state vectors at time step $k$ to be fixed and equal to $\munderbar{\V{x}}_k = \big[ (\V{x}^{(1)}_k)^\top, (\V{x}^{(2)}_k)^\top, \cdots, (\V{x}^{(\no)}_k)^\top \big]^\top\rmv\rmv\rmv\rmv$, the following are assumed to hold for the multiobject and sensor system.    

\begin{description}[labelwidth=\widthof{\ref{ass_unres_mb}}] 
		\item[\namedlabel{ass_dyn}{A.1}] \textbf{Object dynamics.} The object states $\RV{x}^{(1)}_k, \RV{x}^{(2)}_k, \cdots, \RV{x}^{(\no)}_k$ are independent. Furthermore, each object evolves in time according to a known motion model. Specifically, the stochastic process $\{ \RV{x}^{(i)}_k\}_k $ for each object $i$ is Markovian with known conditional density $f^{(i)}_{k+1\vert k}(\V{x}\vert \V{y})$, i.e., the probability of $\RV{x}^{(i)}_{k+1} \in \Set{A}$ conditioned on the event $\{ \RV{x}^{(i)}_{k} = \V{x}^{(i)}_k\}$ is given by $ \int_{\Set{A}} f^{(i)}_{k+1\vert k}(\V{x}\vert \V{x}^{(i)}_k)\dx \V{x}$ for any Borel set $\Set{A}$\vspace{2mm}.

\item[\namedlabel{ass_unres0}{A.2}] \textbf{Resolution graph~\cite{SveUlmHam:12}.} The resolution graph $\RS{G}$ is a random graph spanning the objects ($\Set{V}_{\RS{G}} = \Set{O} $) and with edges sampled independently for each pair $i, l \in \Set{O}$ with\vspace{-.5mm} coupling probability\footnote{Note that other choices for $\Puk$ are possible, for example in~\cite{ChaDun:R82}, $\Puk =1$ for closely spaced objects and gradually decreases to zero for objects that are farther apart than some threshold value.} 
\begin{multline}
   \Puk(\V{x}_k^{(i)}, \V{x}_k^{(l)}) =  \exp \Big( - \frac{1}{2}  \big(\h(\V{x}_k^{(i)}) - \h(\V{x}_k^{(l)}) \big)^{\top} \\[1mm]
     \times \Au^{-1} \big(\h(\V{x}_k^{(i)}) - \h(\V{x}_k^{(l)}) \big) \Big)
    \label{eq_def_Pu}
    \vspace{-2mm} 
\end{multline}
where 
$\h \colon \mathbb{X} \to \mathbb{Z}$ is the sensor measurement function that maps object states $\V{x} \in  \mathbb{X}$ to measurements $\V{z} $ in the measurement domain $\mathbb{Z}$. The sensor resolution capabilities are captured by the positive definite matrix $\Au$. Intuitively, two closely-spaced objects in the measurement domain $\mathbb{Z}$ have a high coupling probability, \ie, high probability of becoming an unresolved pair. Sensor-specific examples of $\h(\cdot)$ and $\Au$ are given in~\cite{KocKeu:J97}. Also denote $\Quk(\V{x}_k^{(i)}, \V{x}_k^{(l)}) \triangleq 1-\Puk(\V{x}_k^{(i)}, \V{x}_k^{(l)})$ for any pair $(i,l)$ of objects. Each component $\RS{C} $ of the resolution graph $\RS{G}$ contains a group of objects. If ${\RS{C}}$ is a singleton, then we call the respective object resolved. Conversely, when $\card{{\RS{C}}} \geq 2$, we refer to the group of objects in ${\RS{C}}$ as unresolved.  
\vspace{1mm} 



\item[\namedlabel{ass_unres_mb}{A.3}] \textbf{Object group detection and measurement.} Given a resolution graph $G$, the group of objects defined by each component of $G$ may generate up to one measurement. For each subset $\Set{I}\subset \Set{O}$, let $\Pdk{\Set{I}} \colon \mathbb{X}^{\card{\Set{I}}} \to [0,1]$ be the probability of detection of group $\Set{I} $ as a function of the object states. Also, let $\fuk{\Set{I}}(\cdot \vert  \munderbar{\V{x}}_k^{\Set{I}})$ be a probability density function on the domain $\mathbb{Z}$. Note that both $\Pdk{\Set{I}}$ and $\fuk{\Set{I}}$, as functions of the object states $(\V{x}^{(i)}\colon i\in \Set{I})$, are symmetric, \ie, the order of object state vectors does not affect the function values. Additionally $\Pdk{\Set{I}}$ and $\fuk{\Set{I}}$ may depend on the group $\Set{I}$, \eg, group size and identity of the unresolved objects. For each component ${C}$ of $G$, the group of objects in $C $ is detected with probability $\Pdk{C}(\munderbar{\V{x}}_k^{C})$, whereupon the group generates a measurement vector $\RV{z} \in \mathbb{Z}$ according to the conditional probability distribution $\fuk{C}(\cdot \vert  \munderbar{\V{x}}_k^{C})$\footnote{For a radar sensor, an unresolved measurement may be generated as $\RV{z} = \sum_{i\in C} w_i \h(\V{x}_k^{(i)}) + \RV{v}$, where $\{w_i\}$ is a convex set of weights (each $w_i$ may depend on the radar cross-section of the $i$-th object) and $\RV{v}$ is an additive noise term that may depend on the group $C$ (e.g., the size of the group).}. No measurement is generated when the group is mis-detected.
\vspace{1mm} 


\item[\namedlabel{ass_clutter}{A.4}] \textbf{Clutter measurements.} Independent of the objects and their corresponding measurements, a random number of clutter measurements \cite{BarLi:B95} are generated as independent and identically distributed (iid) with pdf $\ffak(\cdot)$. The number of such clutter measurements follows a Poisson distribution with rate parameter $\lfak$. We also make the generally-assumed hypothesis that $\ffak(\V{z})>0$ for any $\V{z}\in \mathbb{Z}$.
	
\item[\namedlabel{ass_da}{A.5}] \textbf{Measurement origin uncertainty.} Both clutter and object-originated measurements are arranged in a concatenated measurement vector, $\munderbar{\RV{z}}_k = \big[ (\RV{z}_k^{(1)})^\top, \cdots, (\RV{z}_k^{(\rv{m}_k)})^\top\big]^\top$\rmv\rmv\rmv, with the number of measurements $\rv{m}_k$ being a random variable. The order of measurements in this vector is irrelevant. Furthermore, \emph{measurement origin uncertainty} is assumed, that is, the origin---whether a single object, a group of objects, or clutter---of a single measurement $\RV{z}_k^{(j)}$ is unknown.  	
	
%
%
\end{description}


\section{A Probabilistic Model for Object Partitions}
\label{sec_unresolved_model}
The resolution graph of Assumption \ref{ass_unres0} implicitly introduces a probability measure over the set of all partitions of the object set. More specifically, each edge is a Bernoulli random variable with probability \eqref{eq_def_Pu} and the components of the resulting random graph $\RS{G}$ define a random partition of $\Set{O}$. 

In this section, we construct a probability mass function for partitions of the multiobject set $\Set{O}$, according to the generative model of Assumption \ref{ass_unres0}. Note that here, we extend our previous model from \cite{SauMey:C23} to groups of unresolved objects of arbitrary size.

\subsubsection{Constructive definition}
Consider again the state vectors at time step $k$ to be fixed and equal to $\munderbar{\V{x}}_k = \big[ (\V{x}^{(1)}_k)^\top, (\V{x}^{(2)}_k)^\top, \cdots, (\V{x}^{(\no)}_k)^\top \big]^\top\rmv\rmv\rmv\rmv$. More specifically, any partition $\Set{P} \triangleq \{\Set{U}_1, \Set{U}_2, \cdots, \Set{U}_m\} \in \mathfrak{B}({\Set{O}})$ contains groups of objects such that $\cup_i\, \Set{U}_i = \Set{O}$ and $\Set{U}_i \cap \Set{U}_j = \emptyset$ for all $i\neq j$. 

Sampling a partition $\Set{P}$ from $\mathfrak{B}({\Set{O}})$ can be achieved by first sampling a graph $G$ from the set $\mathfrak{G}({\Set{O}})$ of undirected graphs with vertex set $\Set{O}$. The sampling of $G$, or equivalently of its edge set $\mathcal{E}_{{G}}$, is achieved by independent Bernoulli sampling for each edge $(i,l)$ from the set $ \{ (i,l)\in \intset{1}{\no}^2 \colon i<l\}$ with probability of success given by \eqref{eq_def_Pu}. Second, the set of vertices corresponding to each component of $G$ becomes an element in the partition $\Set{P}$. Sampling such a partition requires $\binom{\no}{2}$ operations, namely one Bernoulli experiment per possible edge. Note that multiple graphs may correspond to a single partition, as showcased in Fig. \ref{figure_graphs}.


\usetikzlibrary{positioning, fit, calc, backgrounds, decorations.pathmorphing, shapes.geometric}
 
\pgfdeclarelayer{background}
\pgfsetlayers{background,main}
 
 
 \tikzset{
  blob/.style={
    thick,
    draw=yellow!80!blue!15,
    fill=yellow!80!blue!15,
    fill opacity=0.6,
    rounded corners=4pt
  }
}
 
\tikzset{
  mystar/.style={
    star,
    star points=5,
    draw=yellow!80,
    fill=yellow!80,
    minimum size=2mm,
    inner sep=0.5pt,
    star point ratio=0.4,  
    rotate=36
  }
}
 
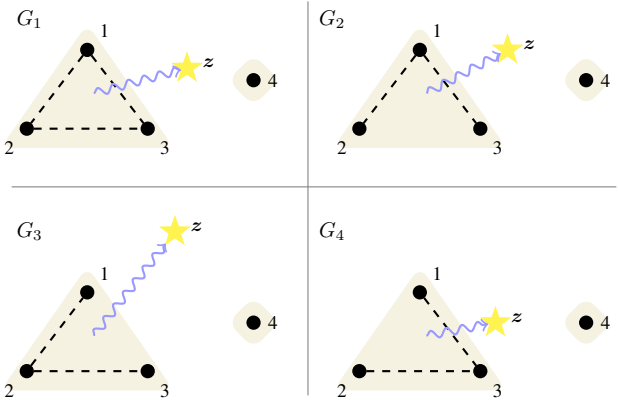
\begin{figure}[t!]
\begin{minipage}[H!]{0.5\textwidth}
\begin{tikzpicture}[
    scale=0.8,                
    transform shape,          
    vertex/.style={circle, draw, fill=black, thick, minimum size=2mm, inner sep=0pt},
    labelstyle/.style={font=\small}
]
\def\xxshift{0.5}
\def\xshift{2.0}
\def\xyshift{1}
\def\xyshiftt{0.7}
\node at (-1.70,2.5) {$G_1$};

\node[mystar] (s1) at (0.9, 1.7) {};  

\node[vertex] (v1) at (-0.75,{2*\xyshift}) {};
\node[vertex] (v2) at (-0.75-\xyshift,\xyshiftt) {};
\node[vertex] (v3) at (-0.75+\xyshift,\xyshiftt) {};
\node[vertex] (v4) at (\xshift,1.5) {};

\draw[thick,dashed] (v1) -- (v2);
\draw[thick,dashed] (v1) -- (v3);
\draw[thick,dashed] (v2) -- (v3);


\node[labelstyle, above right=0.1pt and 0.1pt of v1] {1};
\node[labelstyle, below left=0.1pt and 0.1pt of v2] {2};
\node[labelstyle, below right=0.1pt and 0.1pt of v3] {3};
\node[labelstyle, right=0.1pt of v4] {4};
\node[labelstyle, right=0.5pt of s1] {$\V{z}$};

\coordinate (c1) at ($(v1)!0.33!(v2)!0.33!(v3)$);
\draw[
  ->,
  thick,
  decorate,
  draw=blue!40,
  decoration={
    snake,
    amplitude=1.5pt,
    segment length=6pt
  }
]
($(c1)+(0,0.0)$) -- (s1);

\node at (5-1.7, 2.5) {$G_2$};

\node[mystar] (s2) at (6.2, 2.0) {};  

\node[vertex] (w1) at (-0.75+\xxshift+5,{2*\xyshift}) {};
\node[vertex] (w2) at (-0.75+\xxshift+5-\xyshift,\xyshiftt) {};
\node[vertex] (w3) at (-0.75+\xxshift+5+\xyshift,\xyshiftt) {};
\node[vertex] (w4) at (\xxshift+5+\xshift,1.5) {};

\coordinate (c1) at ($(v1)!0.33!(v2)!0.33!(v3)$);

\draw[thick,dashed] (w1) -- (w2);
\draw[thick,dashed] (w1) -- (w3);


\node[labelstyle, above right=0.1pt and 0.1pt of w1] {1};
\node[labelstyle, below left=0.1pt and 0.1pt of w2] {2};
\node[labelstyle, below right=0.1pt and 0.1pt of w3] {3};
\node[labelstyle, right=0.1pt of w4] {4};
\node[labelstyle, right=0.5pt of s2] {$\V{z}$};

\coordinate (c2) at ($(w1)!0.33!(w2)!0.33!(w3)$);
\draw[
  ->,
  thick,
  decorate,
  draw=blue!40,
  decoration={
    snake,
    amplitude=1.5pt,
    segment length=6pt
  }
]
($(c2)+(0,0.0)$) -- (s2);

\node at (-1.70, -1.0) {$G_3$};

\node[mystar] (s3) at (0.7, -1.0) {}; 

\node[vertex] (u1) at (-0.75,-{2}) {};
\node[vertex] (u2) at (-0.75-\xyshift,-4+\xyshiftt) {};
\node[vertex] (u3) at (-0.75+\xyshift,-4+\xyshiftt) {};
\node[vertex] (u4) at (\xshift,-2.5) {};

\draw[thick,dashed] (u2) -- (u3);
\draw[thick,dashed] (u2) -- (u1);


\node[labelstyle, above right=0.1pt and 0.1pt of u1] {1};
\node[labelstyle, below left=0.1pt and 0.1pt of u2] {2};
\node[labelstyle, below right=0.1pt and 0.1pt of u3] {3};
\node[labelstyle, right=0.1pt of u4] {4};
\node[labelstyle, right=0.5pt of s3] {$\V{z}$};

\coordinate (c3) at ($(u1)!0.33!(u2)!0.33!(u3)$);
\draw[
  ->,
  thick,
  decorate,
  draw=blue!40,
  decoration={
    snake,
    amplitude=1.5pt,
    segment length=6pt
  }
]
($(c3)+(0,0.0)$) -- (s3);

\node at (5-1.7, -1) {$G_4$};

\node[mystar] (s4) at (6.0, -2.5) {};  

\node[vertex] (t1) at (-0.75+\xxshift+5,-2) {};
\node[vertex] (t2) at (-0.75+\xxshift+5-\xyshift,-4+\xyshiftt) {};
\node[vertex] (t3) at (-0.75+\xxshift+5+\xyshift,-4+\xyshiftt) {};
\node[vertex] (t4) at (\xxshift+5+\xshift,-2.5) {};

\draw[thick,dashed] (t1) -- (t3);
\draw[thick,dashed] (t2) -- (t3);


\node[labelstyle, above right=0.1pt and 0.1pt of t1] {1};
\node[labelstyle, below left=0.1pt and 0.1pt of t2] {2};
\node[labelstyle, below right=0.1pt and 0.1pt of t3] {3};
\node[labelstyle, right=0.1pt of t4] {4};
\node[labelstyle, right=0.5pt of s4] {$\V{z}$};

\coordinate (c4) at ($(t1)!0.33!(t2)!0.33!(t3)$);
\draw[
  ->,
  thick,
  decorate,
  draw=blue!40,
  decoration={
    snake,
    amplitude=1.5pt,
    segment length=6pt
  }
]
($(c4)+(0,0.0)$) -- (s4);

\draw[gray] (2.9, 2.8) -- (2.9,-3.7);

\draw[gray] (-2,-0.26) -- (8,-0.26);


\begin{pgfonlayer}{background}
\draw[blob]
  ($(v1)+(0,0.4)$) --
  ($(v2)+(-0.4,-0.3)$) --
  ($(v3)+(0.4,-0.3)$) --
  cycle;
  
  \draw[blob]
  ($(v4)+(0.4,0)$) --
  ($(v4)+(0,0.4)$) --
  ($(v4)+(-0.4,0)$) --
  ($(v4)+(0,-0.4)$) --
  cycle;

\draw[blob]
  ($(w1)+(0,0.4)$) --
  ($(w2)+(-0.4,-0.3)$) --
  ($(w3)+(0.4,-0.3)$) --
  cycle;
  
  \draw[blob]
  ($(w4)+(0.4,0)$) --
  ($(w4)+(0,0.4)$) --
  ($(w4)+(-0.4,0)$) --
  ($(w4)+(0,-0.4)$) --
  cycle;
 
\draw[blob]
  ($(u1)+(0,0.4)$) --
  ($(u2)+(-0.4,-0.3)$) --
  ($(u3)+(0.4,-0.3)$) --
  cycle;
  
  \draw[blob]
  ($(u4)+(0.4,0)$) --
  ($(u4)+(0,0.4)$) --
  ($(u4)+(-0.4,0)$) --
  ($(u4)+(0,-0.4)$) --
  cycle;
 
\draw[blob]
  ($(t1)+(0,0.4)$) --
  ($(t2)+(-0.4,-0.3)$) --
  ($(t3)+(0.4,-0.3)$) --
  cycle;
  
  \draw[blob]
  ($(t4)+(0.4,0)$) --
  ($(t4)+(0,0.4)$) --
  ($(t4)+(-0.4,0)$) --
  ($(t4)+(0,-0.4)$) --
  cycle;

\end{pgfonlayer}

\end{tikzpicture}
\end{minipage}
\caption{Figure showcasing all possible resolution graphs $\{G_i\}_{i=1}^4$ for the partition $\big\{  \{1,2,3\} , \{4\}\big\}$. The graph vertices $\bullet$ are objects with fixed locations $\{\V{x}_k^{(i)}\}_{i=1}^4$. The four graphs have edge sets $\mathcal{E}_{{G_1}} =  \{(1,2), (2,3), (1,3) \} $, $\mathcal{E}_{{G_2}} =  \{(1,2), (1,3) \} $, $\mathcal{E}_{{G_3}} = \{(1,2), (2,3) \}$, and $\mathcal{E}_{{G_4}} = \{(1,3), (2,3) \}$. For each case $i\in \{1,2,3,4\}$ and conditionally on the object set and $G_i$, we additionally depict an instance of a single unresolved measurement $\textcolor{yellow}{\bigstar}$ that is generated by the object group $\Set{I}\triangleq \{1,2,3\}$, i.e., $\V{z}\sim f_k^{\Set{I}}(\cdot \vert \munderbar{\V{x}}^{\Set{I}})$. In all four cases, object $4$ is assumed to be misdetected. Given $\V{z}$, the true association vectors in all four cases are $\V{a}_k = [1, 1, 1, 0]^\top$, $\V{b}_k^{(0)} =[0,0,0,1]^\top$, and $\V{b}_k^{(1)} =[1,1,1,0]^\top$.}
\label{figure_graphs}
\end{figure}

\subsubsection{Probability measure on partitions}
We denote via $\mathcal{G}(\Set{O}, \Puk)$ the distribution of the random graph $\RS{G}$ resulting from the previous construction. This model can be seen as a generalization of the celebrated Erd\H{o}s–R\'{e}nyi random graph model, whose edges are iid with a fixed probability. 

For a given set $\Set{A}\subseteq \Set{O}$, we define 
\begin{equation}
\Vuk{\Set{A}}(\munderbar{\V{x}}^{{\Set{A}}}_k) \triangleq \prod_{\{i,l\} \subset {\Set{A}}} \Quk(\V{x}_k^{(i)}, \V{x}_k^{(l)})\,.
 \end{equation}
 The probability of $\RS{G}$ taking the value ${G}$ is given by\footnote{For all times $k$ and $i\neq j$, we assume that $\V{h}(\V{x}^{(i)}_k ) \neq \V{h}(\V{x}^{(j)}_k )$. In general, events of the type $\big\{\V{h}(\RV{x}_k^{(i)} ) = \V{h}(\RV{x}_k^{(j)} )  \big\}$ have zero probability. Whenever this is not the case, one may eliminate the pairs $(i,j)$ from the edge set.}
\begin{IEEEeqnarray}{rCl}
    \mathbb{P}_{\RS{G}}( {G} \vert \munderbar{\V{x}}_k) &=&  \!\!\! \prod_{(i,l) \in \mathcal{E}_{{G}}} \!\!\!\! \Puk(\V{x}_k^{(i)}, \V{x}_k^{(l)}) \!\! \prod_{(i,l) \notin \mathcal{E}_{{G}}} \!\!\! \Quk(\V{x}_k^{(i)}, \V{x}_k^{(l)}) \nn \\
    &=&  \Vuk{\Set{O}}(\munderbar{\V{x}}_k) \prod_{(i,l) \in \mathcal{E}_{{G}}} \frac{\Puk(\V{x}_k^{(i)}, \V{x}_k^{(l)})}{\Quk(\V{x}_k^{(i)}, \V{x}_k^{(l)})}
 \end{IEEEeqnarray}
for which one can easily check that $\sum_{{G}} \mathbb{P}_{\RS{G}}( {G} \vert \munderbar{\V{x}}_k) = 1$, as a special case of the multi-Binomial theorem.

Subsequently, the random graph $\RS{G}$ induces a random partition $\RS{P} \in \mathfrak{B}({\Set{O}}) $ according to the procedure in the previous section. Thus, the induced probability measure over partitions $\Set{P}\in \mathfrak{B}({\Set{O}})$ is obtained by summing over all graphs with components equal to the elements in $\Set{P}$, i.e., 
\begin{align}
    \mathbb{P}_{\RS{P}}( \Set{P} \vert \munderbar{\V{x}}_k) &= \sum_{G \in \mathfrak{G}({\Set{O}})} \mathbb{P}_{\RS{G}}( {G} \vert \munderbar{\V{x}}_k) \,  \delta_{\Set{C}(G)} (\Set{P}) \label{eq_prob_part_def}
\end{align}
where $\Set{C}(G) $ denotes the set of components of the graph $G$. The computational complexity of constructing $\mathbb{P}_{\RS{P}}$ is gouverned by the enumeration of all $2^{\binom{\no}{2}}$ undirected graphs on the vertex set $\Set{O}$. 

For any subset $\Set{I} \subset \Set{O}$ of objects, let 
\begin{align}
  \ruk{\Set{I}} ( \munderbar{\V{x}}_k^{\Set{I}} ) \triangleq  & 
   \begin{cases}
  \sum_{\substack{G\in \mathfrak{G}({\Set{I}})\\ \Set{C}(G) = \{\Set{I} \}}} \prod_{(i,l) \in \mathcal{E}_G } \frac{\Puk(\V{x}_k^{(i)}, \V{x}_k^{(l)})}{\Quk(\V{x}_k^{(i)}, \V{x}_k^{(l)})}  & \text{ if } \card{\Set{I}} >1\\
   1 & \text{ if } \card{\Set{I}} =1
   \end{cases}\label{eq_qI}
\end{align}
where in equation \eqref{eq_qI}, the summation is carried out over all connected graphs restricted to $\Set{I}$ as the set of vertices. 

\begin{proposition}
\label{prop_ruI}
 For any partition $\Set{P}\in \mathfrak{B}(\Set{O})$, the probability measure $\mathbb{P}_{\RS{P}}( \Set{P} \vert \munderbar{\V{x}}_k)$ can be equivalently expressed as 
\begin{align}
    \mathbb{P}_{\RS{P}}( \Set{P} \vert \munderbar{\V{x}}_k) 
    &= \Vuk{\Set{O}}(\munderbar{\V{x}}_k)  \prod_{\Set{U} \in \Set{P} }\big[ \ruk{\Set{U}} ( \munderbar{\V{x}}_k^{\Set{U}} ) \big]\,.
\end{align} 
\end{proposition} 
Proof of the above proposition is deferred to Appendix \ref{app_prop_ruI}.


\section{Data Association Model and Posterior Distribution}
\label{sec:posterior}

In what follows, we will present the statistical model for data association with potentially unresolved objects.

\subsection{Data Association Variables and Constraints}
Due to the measurement origin uncertainty of Assumption $\ref{ass_da}$, there exist numerous possible explanations for an observed measurement vector $\munderbar{\V{z}}_k = \big[ (\V{z}_k^{(1)})^\top, \cdots, (\V{z}_k^{({m}_k)})^\top\big]^\top\rmv\rmv\rmv$. This fact is represented by the random object-oriented association vector $\RV{a}_{k} = [ \rv{a}_{k}^{(1)}, \cdots\rmv, \rv{a}_{k}^{(\no)} ]^\top$ with entries
\begin{equation}
\rv{a}_{k}^{(i)} \triangleq \begin{cases}
    j \in \intset{1}{m_k} & \text{object $i$ generated/contributed}\\[-.5mm]
    & \text{to the $j$-th measurement} \\[.7mm]
    0 & \text{object $i$ has not contributed} \\[-.5mm]
    & \text{to any measurement.}
\end{cases}
\nonumber
\end{equation}
Note that all vectors $\V{a}_{k} \rmv\in\rmv \Set{A}_k \triangleq \intset{0}{m_k}^{\no}  $ are valid. 

The knowledge contained in $\V{a}_{k}$ explains the origin of every measurement in $\munderbar{\V{z}}_k$. However, $\RV{a}_{k}$ is a random hidden variable in our Bayesian state estimation problem. For minimum mean squared error estimation, nuisance parameters are marginalized by computing a marginal posterior pdf from a joint posterior pdf. Within this marginalization step, for each possible multiobject state vector $ \munderbar{\V{x}}_k$, a sum with a number of terms that is equal to all possible association vectors $\V{a}_{k}$ needs to be computed. Due to the large number of possible vectors, this is typically infeasible.

As in \cite{WilLau:J14,MeyKroWilLauHlaBraWin:J18,ShaSauBucVar:J19,MeyBraWilHla:J17}, we aim to develop a feasible approximate object state estimation based on LBP. Here, the aforementioned infeasible marginalization is avoided by introducing the complementary measurement-oriented association tuple $ \munderbar{\RV{b}}_k = \big({\RV{b}}_k^{(0)}, {\RV{b}}_k^{(1)}, \cdots, {\RV{b}}_k^{(m_k)}   \big)$, where the measurement-oriented association vectors $ \RV{b}_k^{(j)} $, for all $j\in \intset{0}{m_k}$, are defined as 
\begin{align}
\RV{b}_{k}^{(j)} \triangleq \big[ [\RV{b}_{k}^{(j)}]_1, [\RV{b}_{k}^{(j)}]_2, \cdots, [\RV{b}_{k}^{(j)}]_{\no} \big]^\top \in \{0,1\}^{\no} \,. \nonumber 
\end{align}
For $1 \leq j \leq m_k$, the vector $ \RV{b}_k^{(j)} \in \mathbb{B} \triangleq \{0,1\}^{\no}$ indicates the set of objects $\big\{ i\in \intset{1}{\no}\colon [\RV{b}_k^{(j)}]_i = 1 \big\}$ that generated the $j$-th measurement or, whenever $ \V{b}_k^{(j)} = \V{0}_{}$, it indicates that the $j$-th measurement is clutter. The special case of vector $ \RV{b}_k^{(0)}$ indicates miss-detections, \ie, the set $\big\{ i\in \intset{1}{\no}\colon [\RV{b}_k^{(0)}]_i = 1\big\}$ contains all miss-detected objects. An example of association vectors is given in Fig. \ref{figure_graphs}.

Furthermore, the following transformations are defined as
\begin{IEEEeqnarray}{rCl}
    \Set{S}\big(\RV{b}_{k}^{(j)}\big) & \triangleq & \big\{ i \in \intset{1}{\no}\colon [\RV{b}_{k}^{(j)}]_{i} = 1 \big\} \label{eq_set_b} \quad \forall j\in \intset{0}{m_k} \\
    \Set{S}(\munderbar{\RV{b}}_{k}) & \triangleq & \big\{ \Set{S}\big(\RV{b}_{k}^{(j)}\big) \colon j \in \intset{1}{m_k}, \, \Set{S}\big(\RV{b}_{k}^{(j)}\big) \neq \emptyset \big\} \,. \label{eq_set_B}
\end{IEEEeqnarray}

Note that not all elements $ \munderbar{\V{b}}_k \in \Set{B}_k \triangleq \mathbb{B}^{(m_k+1)}$ are valid measurement-oriented association tuples. A $ \munderbar{\V{b}}_k$ is invalid if an object contributes to more than one measurement, or whenever a measurement is assigned to clutter and objects, or whenever an object is both miss-detected and associated to a measurement. A constraint function that eliminates invalid $ \munderbar{\V{b}}_k$ is next constructed by observing that a one-to-one correspondence exists between $\V{a}_{k}$ and valid $ \munderbar{\V{b}}_k$. For example, in the all-clutter case, the measurement-oriented association tuple defined by $\V{b}_{k}^{(0)} =\V{1}$, and $\V{b}_{k}^{(j)} =\V{0}$,  $\forall j\geq 1$, uniquely corresponds to $\V{a}_{k} =[0, \cdots\rmv, 0]^\top\rmv\rmv$. It can easily be verified that for invalid $\munderbar{\V{b}}_{k}$ there is no corresponding valid $\V{a}_{k}$. For example, an invalid event where a single object contributes to two different measurements can be described by an $\munderbar{\V{b}}_k \in  \Set{B}_k$ but not by an $\V{a}_k \in \Set{A}_k$.

Eliminating non-valid measurement-oriented association tuples can be achieved by checking correspondence between events $\V{a}_{k}$ and $ \munderbar{\V{b}}_k$. This correspondence is enforced by the following product of elementwise consistency constraints, i.e.,
\begin{equation}
    \Psi_k (\V{a}_k, \munderbar{\V{b}}_k) \ist \triangleq \ist \prod_{i=1}^{\no} \prod_{j=0}^{m_k} \ist\ist \psi_{ij}\big({a}_k^{(i)}\rmv\rmv, \V{b}_k^{(j)}\big) \label{eq_psi}
\end{equation}
where for all $(i,j)\in \intset{1}{\no} \times \intset{0}{m_k}$, we define
\makeatletter
\newcommand\niton{\mathrel{\m@th\mathpalette\canc@l\owns}}
\newcommand\canc@l[2]{{\ooalign{$\hfil#1/\mkern1mu\hfil$\crcr$#1#2$}}}
\makeatother
\begin{align}
\label{eq_psi_ij}
{ 
    \psi_{ij}({a}_k^{(i)}\rmv\rmv, \V{b}_k^{(j)}) \triangleq \begin{cases}
        0, & a_k^{(i)}=j \text{ and } i \notin \Set{S}\big( \V{b}_{k}^{(j)} \big), \text{ or} \\[.5mm] 
        &    a_k^{(i)} \neq j \text{ and } i \in \Set{S}\big( \V{b}_{k}^{(j)} \big)  \\[1.5mm]
        1, & \text{otherwise}\,.
    \end{cases}}
\end{align}
Only if all elements in $\V{a}_{k}$ are consistent with all elements in $\V{b}_{k}$, i.e., all element-wise consistency constraints in \eqref{eq_psi} are equal to one, do the two vectors represent the same association event. Since only valid events can be represented by an $\V{a}_{k}$ and a $\munderbar{\V{b}}_{k}$ at the same time, the represented event must be valid.

\subsection{A Model for the Prior Association Probabilities}

A probabilistic model for the association vectors $\V{a}_k \in \Set{A}_k$ and $\munderbar{\V{b}}_k \in \Set{B}_k$ arises due to the mechanism of generating and detecting measurements as described in Section \ref{sec:model}. This model serves as a prior distribution for the aforementioned association variables and is given by the following proposition.

\begin{proposition}
\label{prop_prior1_aB}
The prior pmf of $\RV{a}_k \in \Set{A}_k$, $\munderbar{\RV{b}}_k \in \Set{B}_k$, and $\rv{m}_k$ is 
\begin{align}
     p(\V{a}_k, \munderbar{\V{b}}_k ,  m_k \vert \munderbar{\V{x}}_{k}) &= \Psi_k (\V{a}_k, \munderbar{\V{b}}_k) \frac{e^{-\lfa} (\lfa)^{m_k-\card{\Set{S}({\munderbar{\V{b}}_k})}}}{m_k!} \nn \\
& \times  \Vuk{\Set{O}}(\munderbar{\V{x}}_k) \Bigg[  \prod_{\Set{A} \in \Set{S}(\munderbar{\V{b}}_k)} \Pdk{\Set{A}}(\munderbar{\V{x}}_k^{\Set{A}})\,  \ruk{\Set{A}} (\munderbar{\V{x}}_k^{\Set{A}} ) \Bigg] \nonumber \\ 
&  \times  \!\!\!\!\! \!\!\!\!\! \sum_{\Set{U} \in \mathfrak{B} (\Set{S}(\V{b}_k^{(0)})) }  \prod_{\Set{W}\in \Set{U}}  \big( 1- \Pdk{\Set{W}}(\munderbar{\V{x}}_k^{\Set{W}})\big) \,  \ruk{\Set{W}} (\munderbar{\V{x}}_k^{\Set{W}} )  \label{eq_prior_a2}
\end{align}
where $\Set{S}(\V{b}_k^{(0)})  $ is the set of misdetected objects according to $\munderbar{\V{b}}_k$, while $\Pdk{\hspace{.15mm} \Set{A}}(\munderbar{\V{x}}_k^{\Set{A}})$ is the probability of detection for group $\Set{A}$, introduced in Sec. \ref{sec:model}, Assumption A.3.

\end{proposition}
The derivation of Proposition \ref{prop_prior1_aB}  is based on Proposition \ref{prop_ruI} and is provided in Appendix \ref{app_prior_a}.

\subsection{The Joint Posterior Pdf}
\vspace{-.5mm}

The goal of this paper is to compute the marginal probabilities $f(\V{x}_k^{(i)}\vert \munderbar{\V{z}}_{1:k})$, for each object $i\in \mathcal{O}$. In principle, these marginal pdfs could be obtained from the joint posterior pdf $f(\munderbar{\V{x}}_{k}, \V{a}_k, \munderbar{\V{b}}_k \vert \munderbar{\V{z}}_{1:k})$, which involves all object states and association variables at time $k$, by direct marginalization. However, this is typically infeasible due to the large number of valid association events. To obtain feasible estimates of the required marginals, we apply LBP on the factor graph representing $f(\munderbar{\V{x}}_{k}, \V{a}_k, \munderbar{\V{b}}_k \vert \munderbar{\V{z}}_{1:k})$ \cite{KscFreLoe:01}.

In what follows, we assume that the measurement at time $k$, $\V{z}_{k}$, has been observed and is thus fixed. Note that this implies that $m_k$ is also observed and fixed. By using Bayes' rule, the joint posterior factorizes as
\begin{IEEEeqnarray}{rCl}
    \hspace{0mm}f(\munderbar{\V{x}}_{k}, \V{a}_k, \munderbar{\V{b}}_k \vert \munderbar{\V{z}}_{1:k}) &=& f(\munderbar{\V{x}}_{k}, \V{a}_k, \munderbar{\V{b}}_k \vert \munderbar{\V{z}}_{1:k-1}, \munderbar{\V{z}}_k, m_k) \nonumber \\[2mm]
    & \propto& f( \munderbar{\V{z}}_k \vert \munderbar{\V{x}}_{k}, \V{a}_k, \munderbar{\V{b}}_k, m_k) \hspace{.3mm} p(\V{a}_k, \munderbar{\V{b}}_k , m_k\vert \munderbar{\V{x}}_{k}) \nonumber \\[2mm]
    & & \hspace{8mm}   \times f(\munderbar{\V{x}}_{k} \vert \munderbar{\V{z}}_{1:k-1}). \label{eq_full_post}
\end{IEEEeqnarray}
Note that here we have made the common assumption \cite{MeyKroWilLauHlaBraWin:J18} that, conditioned on $\munderbar{\V{x}}_{k}$ and $m_k$, the association vectors $\V{a}_k$ and $\munderbar{\V{b}}_k$ are independent of all previous measurements $\munderbar{\V{z}}_{1:k-1}$.

The first term in \eqref{eq_full_post} corresponds to the likelihood of the measurement $\munderbar{\V{z}}_k$ that, based on the assumptions discussed in Section \ref{sec:model}, can be written\vspace{1mm} as
\begin{align}
   f( \munderbar{\V{z}}_k \vert \munderbar{\V{x}}_{k}, \V{a}_k, \munderbar{\V{b}}_k, m_k)     &=  \prod_{j=1}^{m_k} \ffa\big({\V{z}}_k^{(j)}\big)  g\big(\munderbar{\V{x}}_{k}, \Set{S}(\V{b}_k^{(j)});\V{z}_k^{(j)}\big) \nn\\[-1.5mm]
    \label{eq_full_lik}\\[-8mm]
    \nn
\end{align}
where, for any $j\in \intset{1}{m_k}$ and subset $\Set{I}\subseteq \Set{O}$, we introduced
\begin{equation}
g\big(\munderbar{\V{x}}_{k}, \Set{I};\V{z}_k^{(j)}\big) \triangleq \begin{cases}
    \frac{\fuk{\Set{I}}(\V{z}_k^{(j)}\vert \munderbar{\V{x}}_{k}^{\Set{I}})}{\ffa (\V{z}^{(j)}_k)}, & \text{if }  \Set{I} \neq \emptyset \\
    1, & \text{if } \Set{I}=  \emptyset\,.
\end{cases}
\nonumber
\end{equation}

By plugging expressions \eqref{eq_prior_a2} and \eqref{eq_full_lik} into \eqref{eq_full_post}, and dropping constant factors, we obtain 
\begin{align}
   &f( \munderbar{\V{x}}_{k}, \V{a}_k,  \munderbar{\V{b}}_k \vert \munderbar{\V{z}}_{1:k}) \propto  \Psi_k (\V{a}_k, \munderbar{\V{b}}_k) \Bigg[ \prod_{i=1}^{\no} f\big(\V{x}_k^{(i)} \big\vert \munderbar{\V{z}}_{1:k-1}\big) \Bigg] \nn\\[.5mm]
   &\hspace{8mm} \times  \Bigg[ \prod_{j=0}^{m_k} q_k^{(j)}\big( \munderbar{\V{x}}^{ }_{k}, \Set{S}(\V{b}_k^{(j)})  \big)  \Bigg]  \label{eq_full_post3}
\end{align}
where for any subset $\Set{I} \subset \Set{O}$, we introduce
\begin{align}
& q_k^{(0)}\big(\munderbar{\V{x}}^{ }_{k}, \Set{I}^{ }  \big)  \triangleq \Vuk{\Set{O}}(\munderbar{\V{x}}_k)  \sum_{\Set{P} \in \mathfrak{B}({\Set{I})^{ }}}  \prod_{\Set{U} \in\Set{P}} \big( 1- \Pdk{\hspace{.3mm}\Set{U}}(\munderbar{\V{x}}_k^{\Set{U}})\big) \,  \ruk{\Set{U}} (\munderbar{\V{x}}_k^{\Set{U}} )   \label{eq_q0_generic_def}
\end{align}
while for $j\in \intset{1}{m_k}$
\begin{align}
& q_k^{(j)}\big(\munderbar{\V{x}}^{ }_{k}, \Set{I}  \big)  \triangleq  \begin{cases}
      \frac{ \ruk{\Set{I}} ( \munderbar{\V{x}}_k^{\Set{I}} ) \, \Pdk{\Set{I}}(\munderbar{\V{x}}_k^{\Set{I}}) \, \fuk{\Set{I}} (\V{z}_k^{(j)} \vert \munderbar{\V{x}}_k^{\Set{I}} ) }{\lfa \ffa (\V{z}_k^{(j)})}, & \text{if } {\Set{I} \neq \emptyset} \\[2mm]
    1, & \text{if } {\Set{I} = \emptyset}. \nonumber 
\end{cases}
\end{align}
The factorization of $f(\munderbar{\V{x}}_{k}, \V{a}_k, \munderbar{\V{b}}_k \vert \munderbar{\V{z}}_{1:k})$ in \eqref{eq_full_post3} is key for the development of a scalable LBP method presented next.

\begin{figure}[t!]
\begin{minipage}[H!]{0.5\textwidth}
\centering\hspace{4.5mm}
\vspace{-1mm}
\centering
  \begin{picture}(1,0.45)%
    \lineheight{1}%
    \setlength\tabcolsep{0pt}%
    \put(0,0){\includegraphics[width=\unitlength]{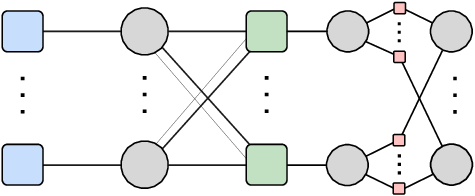}}%
    \put(0.8,-.03){\color[rgb]{0,0,0}\makebox(0,0)[lt]{\lineheight{1.25}\smash{\begin{tabular}[t]{l}$\psi_{m_k,n_{\mathrm{o}}}$\end{tabular}}}}%
    \put(0.80067877,0.433){\color[rgb]{0,0,0}\makebox(0,0)[lt]{\lineheight{1.25}\smash{\begin{tabular}[t]{l}$\psi_{0,1}$\end{tabular}}}}%
    \put(0.77,0.257){\color[rgb]{0,0,0}\makebox(0,0)[lt]{\lineheight{1.25}\smash{\begin{tabular}[t]{l}$\psi_{0,n_{\mathrm{o}}}$\end{tabular}}}}%
    \put(0.75,0.155){\color[rgb]{0,0,0}\makebox(0,0)[lt]{\lineheight{1.25}\smash{\begin{tabular}[t]{l}$\psi_{m_k,1}$\end{tabular}}}}%
    \put(0.935,0.335){\color[rgb]{0,0,0}\makebox(0,0)[lt]{\lineheight{1.25}\smash{\begin{tabular}[t]{l}$\V{a}^1$\end{tabular}}}}%
    \put(0.925,0.054){\color[rgb]{0,0,0}\makebox(0,0)[lt]{\lineheight{1.25}\smash{\begin{tabular}[t]{l}$\V{a}^{n_{\mathrm{o}}}$\end{tabular}}}}%
    \put(0.72,0.333){\color[rgb]{0,0,0}\makebox(0,0)[lt]{\lineheight{1.25}\smash{\begin{tabular}[t]{l}$\V{b}^0$\end{tabular}}}}%
    \put(0.705,0.05){\color[rgb]{0,0,0}\makebox(0,0)[lt]{\lineheight{1.25}\smash{\begin{tabular}[t]{l}$\V{b}^{m_k}$\end{tabular}}}}%
    \put(0.55,0.338){\color[rgb]{0,0,0}\makebox(0,0)[lt]{\lineheight{1.25}\smash{\begin{tabular}[t]{l}\hspace{-0.28mm}$q^0$\end{tabular}}}}%
    \put(0.535,0.055){\color[rgb]{0,0,0}\makebox(0,0)[lt]{\lineheight{1.25}\smash{\begin{tabular}[t]{l}\hspace{-0.34mm}$q^{m_k}$\end{tabular}}}}%
    \put(0.28,0.055){\color[rgb]{0,0,0}\makebox(0,0)[lt]{\lineheight{1.25}\smash{\begin{tabular}[t]{l}$\V{x}^{n_{\mathrm{o}}}$\end{tabular}}}}%
    \put(0.03,0.335){\color[rgb]{0,0,0}\makebox(0,0)[lt]{\lineheight{1.25}\smash{\begin{tabular}[t]{l}$f^1$\end{tabular}}}}%
    \put(0.022,0.055){\color[rgb]{0,0,0}\makebox(0,0)[lt]{\lineheight{1.25}\smash{\begin{tabular}[t]{l}$f^{n_{\mathrm{o}}}$\end{tabular}}}}%
    \put(0.287,0.337){\color[rgb]{0,0,0}\makebox(0,0)[lt]{\lineheight{1.25}\smash{\begin{tabular}[t]{l}$\V{x}^1$\end{tabular}}}}%
  \end{picture}%
\end{minipage}
\vspace{3mm}
\caption{Factor graph showcasing the dependencies between object state variables and association variables at a single time\vspace{.3mm} step $k$. The following compact notation was used: $f^{i} \triangleq f\big(\V{x}_k^{(i)} \big\vert \munderbar{\V{z}}_{1:k-1}\big)$, $q^j \triangleq q_k^{(j)}\big( \munderbar{\V{x}}^{ }_{k}, \Set{S}(\V{b}_k^{(j)})  $, for all $j\in \intset{0}{m_k}$, and $\psi_{ij}\vspace{.5mm} \triangleq \psi_{ij}\big({a}_k^{(i)}\rmv\rmv, \V{b}_k^{(j)}\big)$, all $j\in \intset{1}{m_k}$ and $i\in \intset{1}{\no}$. }
\label{fig_factor_graph}
\end{figure}

\section{A Generic LBP Algorithm} \label{sec_prop_alg1}

In the following, the subscript $k$, denoting the time index, will be dropped for compactness. Furthermore, the time-predicted distribution ${f}\big(\V{x}_k^{(i)} \big\vert \munderbar{\V{z}}_{1:k-1}\big)$ is denoted as ${f}_{0}^{(i)}\big(\V{x}^{(i)} \big)$. The algorithm is composed of a succession of message exchanges between the various variables and factors of \eqref{eq_full_post3}. The succession of these messages is presented next.  

\newcommand{\mxq}[1]{\beta_{\V{x}^i q^{j}}^{(#1)}}
\newcommand{\pah}[3]{\nu_{a^{#1} h_{#2}}^{(#3)}}
\newcommand{\fxh}[3]{\alpha_{\V{x}^{#1} h_{#2}}^{(#3)}}
\newcommand{\muhx}[3]{\gamma_{h_{#1}\V{x}^{#2}}^{(#3)}}

\begin{enumerate}
\item \textbf{Measurement evaluation}. The messages from the joint node $\munderbar{\V{x}}_k$ to each node $\V{b}^{(j)}$, with $j\in \intset{1}{m_k}$, are
\begin{align}
    \varphi_{\munderbar{\V{x}} \V{b}^{j}}(\V{b}) = \int & \frac{ \ru{\Set{I}} ( \munderbar{\V{x}}^{\Set{S}(\V{b})} ) \, \Pd{\Set{S}(\V{b})}(\munderbar{\V{x}}^{\Set{S}(\V{b})}) \, \fu{\Set{I}} (\V{z}^{(j)} \vert \munderbar{\V{x}}^{\Set{S}(\V{b})} ) }{\lfa \ffa (\V{z}^{(j)})} \nn\\[1.5mm]
     & \times \prod_{i \in \Set{S}(\V{b}^{})}^{} {f}_{0}^{(i)}\big(\V{x}^{(i)} \big)   \mathrm{d} \munderbar{\V{x}}^{\Set{S}(\V{b})} \label{eq_meas_eval}
\end{align}
for $\V{b}$ such that $\Set{S}(\V{b})\neq \emptyset$, while $\varphi_{\munderbar{\V{x}} \V{b}^{j}}(\V{b}^{}) = 1 $ for $\Set{S}(\V{b}^{}) = \emptyset$. The message to $\V{b}^{(0)}$ is  
\begin{equation}
\varphi_{\munderbar{\V{x}} \V{b}^{0}}(\V{b}) = \int {q}^{(0)}\big( \munderbar{\V{x}}^{ }, \Set{S}(\V{b}^{})\big) \prod_{i =1}^{\no} {f}_{0}^{(i)}\big(\V{x}^{(i)} \big)   \mathrm{d} \munderbar{\V{x}}^{ } \,.    
\end{equation}

\item \textbf{Data association loop.} Execute for each iteration $\ell = 1,2, \cdots, L$ or until convergence: 
\begin{itemize}
    \item Compute messages $\mu^{(\ell)}_{\V{b}^j a^i}(\cdot)$ from each node $\V{b}_k^{(j)}$ to each node ${a}_k^{(i)}$ (\ie, $\forall (i,j) \in \intset{1}{\no} \times \intset{0}{m}$) via
\begin{align}
    &\mu^{(\ell)}_{\V{b}^j a^i}(a) \propto 
    &\sum_{\V{b}\in \mathbb{B}} \psi_{ij}({a}, \V{b}) \varphi_{\munderbar{\V{x}} \V{b}^{j}}(\V{b})   \prod_{i' \neq i} \mu^{(\ell-1)}_{ a^{i'}  \V{b}^j}(\V{b}) \nonumber
\end{align}
where the initialization $\mu^{(0)}_{ a^{i}  \V{b}^j} = 1$ is used. The above message can be observed to take only two distinct values---for $a = j$ and for $a\neq j$. By dividing $\mu^{(\ell)}_{\V{b}^j a^i}(\cdot)$ with one of the two distinct values, \eg, $\mu^{(\ell)}_{\V{b}^j a^i}(a)$ with $a\neq j$, and performing an equivalent normalization step for the messages $ \mu^{(\ell-1)}_{ a^{i'}  \V{b}^j}(\cdot)$, a simplified form can be obtained as
\begin{equation}
    \mu^{(\ell)}_{\V{b}^j a^i}(a) = \nonumber \\ 
    \begin{cases}
       \Mbat{\ell}_{ij} & \text{ if } a = j \\
       1 & \text{ otherwise }
    \end{cases}
    \label{eq_ba_cases}
\end{equation}
where
\begin{align}
    \Mbat{\ell}_{ij} \triangleq \frac{\sum_{\V{b} \in \mathbb{B}} b_i \ist\ist \varphi_{\munderbar{\V{x}} \V{b}^{j}}\rmv(\V{b}) \, \prod_{i' \neq i} \big[ \Mabt{\ell-1}_{i'j} \big]^{b_{i'}} }{\sum_{\V{b} \in \mathbb{B}} (1-b_i) \ist\ist \varphi_{\munderbar{\V{x}} \V{b}^{j}}\rmv(\V{b}) \, \prod_{i' \neq i} \big[ \Mabt{\ell-1}_{i'j} \big]^{b_{i'}}} \nn \\[.5mm]
     \label{eq_Uba} \\[-7.5mm]
     \nn
\end{align}
The equivalent normalization step for the messages $ \mu^{(\ell-1)}_{ a^{i'}  \V{b}^j}(\cdot)$ resulting in the  expressions $\Mabt{\ell-1}_{ij}$, will be introduced in what follows\vspace{.5mm}.

    \item Compute messages $\mu^{(\ell)}_{a^i \V{b}^j }(\cdot)$ from each node ${a}_k^{(i)}$ to each node $\V{b}_k^{(j)}$ (\ie, $\forall (i,j) \in \intset{1}{\no} \times \intset{0}{m}$) via
    \begin{equation}
    \mu^{(\ell)}_{ a^{i} \V{b}^j}(\V{b})  \propto \sum_{{a}=0}^{m}  \psi_{ij}({a}, \V{b}) \prod_{j' \neq j} \mu^{(\ell)}_{\V{b}^{j'} a^i}(a)  \nonumber
\end{equation}
which, after renormalization, has the form 
\begin{equation}
    \mu^{(\ell)}_{a^i\V{b}^j}(\V{b}) =  \\ 
    \begin{cases}
       \Mabt{\ell}_{ij} & \text{ if } i \in \Set{S}(\V{b}) \\
       1 & \text{ otherwise }
    \end{cases}
    \label{eq_ab_cases}
\end{equation}
where
\begin{equation}
     \Mabt{\ell}_{ij}  \triangleq \frac{ 1   }{\sum_{\substack{j'=0 \\ j'\neq j}}^{m}  \Mbat{\ell}_{ij'} } \nn \,.
\end{equation}
\end{itemize}

\item \textbf{Compute association beliefs (optional).} For all $j\in \intset{0}{m}$ and $i\in \intset{1}{\no}$, compute the probability estimates
\begin{align}
    \hat{p}_{j}(\V{b}) &\propto \varphi_{\munderbar{\V{x}} \V{b}^{j}}(\V{b}) \prod_{i=1}^{\no} \mu^{(L)}_{ a^{i} \V{b}^j}(\V{b}) \nn \\
    \hat{p}_{i}(a) &\propto  \prod_{j=1}^{m} \mu^{(L)}_{ \V{b}^j a^{i}}(a) \nn
\end{align}
which take the following forms 
\begin{equation*}
    \hat{p}^{}_{j}(\V{b}) \propto \begin{cases}
         \varphi_{\munderbar{\V{x}} \V{b}^{j}}(\V{b}) \prod_{i \in \Set{S}(\V{b})}  \Mabt{L}_{ij}  & \text{ if } \Set{S}(\V{b}) \neq \emptyset \\
            1 & \text{ otherwise }
    \end{cases}
\end{equation*}
and 
\begin{equation}
    \hat{p}^{}_{i}(a) \propto \begin{cases}
           \Mbat{L}_{ia}  & \text{ if } a \in \intset{1}{m} \\
            1 & \text{ if } a = 0\,.
    \end{cases} \label{eq_pa_est}
\end{equation}

\item \textbf{Message from each node $\V{q}^j$ to each node $\V{x}^{(i)}$. }For each index $j$, the message from a node $\V{b}^{(j)}$ to the corresponding factor ${q}^j$ is given by
\begin{equation*}
    \mu_{\V{b}^{j}  {q}^j}(\V{b} ) = \prod_{i=1}^{\no} \mu^{(L)}_{ a^{i} \V{b}^j}(\V{b} )
\end{equation*}
where, for each pair $(i,j)$, the message $\mu_{ a^{i}  \V{b}^j}(\V{b})$ takes two distinct values based on $\V{b}$, as seen in \eqref{eq_ab_cases}. Subsequently for $j\in \intset{1}{m}$, the message from $q^j$ to $\V{x}^{(i)}$ becomes 
\begin{IEEEeqnarray}{rCll}
\hspace{-6mm} \gamma_{{q}^j  \V{x}^{i} }(\V{x}^{(i)})   & =  & \sum_{ \substack{\V{b}\in \mathbb{B} } }  \mu_{\V{b}^{j}  {q}^j}(\V{b}) \Big[ &(1-b_i) \, \varphi_{\munderbar{\V{x}} \V{b}^{j}}(\V{b})   \nn\\
 && & +\,  b_i\, \xi_i^{(j)}(\V{x}^{(i)}, \Set{S}(\V{b})) \Big] \label{eq_gamma_qj}
\end{IEEEeqnarray}
where we introduce the marginal likelihood of object $i$, when group $\Set{I} \supseteq \{i\}$ is associated with the $j$-th measurement, as 
\begin{align}
    &\xi_i^{(j)}(\V{x}^{(i)}, \Set{I})  =\nn  \\[2.5mm]
   &\int  \frac{ \ru{\Set{I}} ( \munderbar{\V{x}}^{\Set{I}} ) \, \Pd{\Set{I}}(\munderbar{\V{x}}^{\Set{I}} ) \, \fu{\Set{I}} (\V{z}^{(j)} \vert \munderbar{\V{x}}^{\Set{I} } ) }{\lfa \ffa (\V{z}^{(j)})} \prod_{l \in \Set{I}\setminus i } {f}_{0}^{(l)}\big(\V{x}^{(l)} \big)   \mathrm{d} \munderbar{\V{x}}^{\Set{I}\setminus i}. \nn\\[-3mm]
\end{align}
For the special case of $j=0$, the message from $q^0$ to $\V{x}^{(i)}$ becomes
\begin{IEEEeqnarray}{rCl}
    \gamma_{{q}^0  \V{x}^{i} }(\V{x}^{(i)}) &=& \sum_{\V{b} \in \mathbb{B} } \mu_{\V{b}^{0} q^0}(\V{b}) \, \xi_i^{(0)}(\V{x}^{(i)}, \Set{S}(\V{b})) 
\end{IEEEeqnarray}    
    where 
    \begin{IEEEeqnarray}{rCl}
   \xi_i^{(0)}(\V{x}^{(i)}, \Set{I} ) & =  & \int {q}^{(0)}(\munderbar{\V{x}}^{ }, \Set{I}  ) \prod_{\substack{l \in \Set{O} \setminus i }}^{} f_0^{(l)} (\V{x}^{(l)} ) \mathrm{d} \munderbar{\V{x}}^{\Set{O}\setminus i} \,. \nn
\end{IEEEeqnarray}

\item  \textbf{Object state update.} The belief computed at node $\V{x}^{(i)}$, for all $i$, is given by  
\begin{equation}
    {f}_{ }(\V{x}^{(i)} \vert \munderbar{\V{z}})   \propto  {f}_{0}^{(i)}(\V{x}^{(i)} )  \Bigg[ \prod_{ {j=0}}^{m}   \gamma_{{q}^j  \V{x}^{i} }(\V{x}^{(i)}) \Bigg] \,. \label{eq_belif_x}
\end{equation}

The right-hand side terms in \eqref{eq_belif_x} have the following interpretation: the first term is the prior density for $\V{x}^{(i)}$ while the product over indices $j\in \intset{0}{m}$ corresponds to measurement update terms with $j=0$ being the special case of misdetection. 
\end{enumerate}

Various implementations of the previous generic LBP algorithm are possible. The most flexible choice relies on particle filter methods~\cite[Ch. 3]{RisAruGor:B04}, where the messages are represented via collections of weighted samples. Such implementations incorporate the exact non-linear and/or non-Gaussian object state-space model, however, a curse-of-dimensionality phenomenon occurs whenever multiobject integrals, such as the one in \eqref{eq_meas_eval}, need to be evaluated. Here, the number of samples generally must increase exponentially with the integration dimension in order to maintain evaluation accuracy.  

In this work, we focus on Gaussian implementations, where messages are represented via Gaussian densities or Gaussian mixtures. In contrast to the sampling-based implementations, Gaussian implementations enjoy analytical forms even for high-dimensional integration. However, linearization or other approximations are necessary for Gaussian implementations when the dynamics of objects or the sensor measurement equation are non-linear.    

\subsection{Computational Complexity}
\label{sec_comp_complexity}
The computational complexity of the LBP algorithm heavily depends on the representation of the individual messages. The measurement evaluation step of \eqref{eq_meas_eval} has to be conducted $m\, 2^{\no}$ times. 
An additional complexity in the measurement evaluation of \eqref{eq_meas_eval} stems from the need to integrate over the function $\ru{\Set{I}} ( \munderbar{\V{x}}^{\Set{S}(\V{b})} )$ defined in \eqref{eq_qI} for any subset $\Set{I}\subset \Set{O}$. Evaluating $\ru{\Set{I}}(\cdot)$ at one point involves an additional worst-case complexity of $O(2^{\no ^2})$ which has to be accounted for in sampling-based implementations when evaluating \eqref{eq_meas_eval}. In Gaussian implementations however, the functions $\ru{\Set{I}}(\cdot)$ are assumed (or approximated) as constants as described in the next sections. 

The computational complexity of the data association loop is dominated by the evaluation of the message in \eqref{eq_Uba}, and is $O(m\,\no\, 2^{\no})$. In contrast, a direct approach that computes marginal association probabilities for each object, would involve summations over all association vectors $\V{a}\in \Set{A}$, leading to a computational complexity of the order $O(m^{\no})$. Note that the direct approach also needs to evaluate measurements according to \eqref{eq_meas_eval} and hence also incurs the same complexity as the LBP for measurement evaluation. Furthermore, a computational cost that depends on $2^{\no}$ seems to be unavoidable for any algorithm since all subsets of $\Set{O}$ need to be evaluated.


\section{Gaussian LBP Algorithm} \label{sec_prop_alg2}

In addition to Assumptions \ref{ass_dyn}-\ref{ass_da}, the following assumptions will prove useful in the current section. 

\begin{description}[labelwidth=\widthof{\ref{ass_Pu_constA}}] 
		\item[\namedlabel{ass_dyn_g}{B.1}] \textbf{Gaussian object models.} The prior object densities are Gaussian and the object dynamics are linear-Gaussian. Thus, the predicted object densities are also Gaussian, denoted via $f_{0}^{(i)}(\V{x}) = \fg(\V{x}, \V{m}^{(i)}, \M{P}^{(i)})$ for all $i\in \Set{O}$.
		\item[\namedlabel{ass_Pd_g}{B.2}] \textbf{Constant probabilities of detection.} The probabilities of detection are constants, \ie, for any set $\Set{I}\subseteq \Set{O}$ $\Pdk{\Set{I}}(\munderbar{\V{x}}^{\Set{I}}) = \Pdk{\Set{I}}$ for all $\munderbar{\V{x}}^{\Set{I}}$.
                 \item[\namedlabel{ass_obs_g}{B.3}] \textbf{Gaussian observation models.} The sensor observation functions are linear and the sensor noise is additive Gaussian. Specifically, for any nonempty subset of unresolved objects $\Set{A}  \subset \Set{O}$, the sensor likelihood function is\footnote{Another possibility for the likelihood function is $\fg \big(\V{z}; \sum_{\ell \in \Set{A}} \M{H}_{k, \ell}^{\Set{A}} \, \V{x}_k^{(\ell)}, \M{R}_{k}^{\Set{A}} \big)$, where the observation matrix $\M{H}_{k, \ell}^{\Set{A}}$ depends on both the object group $\Set{A}$ and the individual object index $\ell$. }
\begin{equation}
                 \fuk{\Set{A}}(\V{z} \vert  \munderbar{\V{x}}_k^{\Set{A}}) = \fg \Big(\V{z}; \sum_{\ell \in \Set{A}} \M{H}_{k}^{\Set{A}} \, \V{x}_k^{(\ell)}, \M{R}_{k}^{\Set{A}} \Big)
                 \label{eq_gobs_model}
\end{equation}
where the observation $\M{H}_{k}^{\Set{A}}$ and the covariance $\M{R}_k^{\Set{A}}$ matrices may depend on the unresolved group identity $\Set{A}$. 
Note that this simplified Gaussian model is flexible enough to model sensing modalities where an unresolved measurement correspond to a centroid of the individual object measurements and where the measurement noise varies with the size of the unresolved group.

 \item[\namedlabel{ass_Pu_constA}{B.4a)}] \textbf{Coupling probability based on predicted object states.} The coupling probability of an unresolved pair $\Puk(\cdot, \cdot) $ is a constant $\tPuk$ given by  
 \begin{align}
 \hspace{-.3cm}  \tPuk^{il} = &  \exp \Big( - \frac{1}{2}  \big(\hv^{(i)} - \hv^{(l)} \big)^{\top} \Au^{-1} \big(\hv^{(i)} - \hv^{(l)} \big) \Big)
    \label{eq_def_tPuA} 
\end{align}
    where $\hv^{(i)} \triangleq   \M{H}_{k}^ {\{i\}}\, \V{m}^{(i)}$ for each $i\in \Set{O}$.
    


 \item[\namedlabel{ass_Pu_constB}{B.4b)}] \textbf{Averaged coupling probability.} The coupling probability of an unresolved pair $\Puk(\cdot, \cdot) $ is a constant $\tPuk$ given by  
 \begin{align}
   \tPuk^{il} = & \iint \Puk(\V{x}, \V{y}) f_{k\vert k-1}^{(i)}(\V{x}) f_{k\vert k-1}^{(l)}(\V{y}) \dx \V{x} \dx \V{y}\,.
    \label{eq_def_tPuB}
\end{align}
Note that when combined with assumptions \ref{ass_dyn_g} and \ref{ass_obs_g}, the above simplifies to
 \begin{align}
  \hspace{-.3cm}   \tPuk^{il} =    \sqrt{\text{det}(2\pi \Au)} \, \fg \big(\M{H}_k^ {\{i\}} \V{m}^{(i)} ; \M{H}_k^ {\{l\}} \V{m}^{(l)}, \M{C}^{il}_k  \big)\,.
    \label{eq_def_tPuB2}
\end{align}
 where $\M{C}^{il}_k = \Au + \M{H}_k^ {\{i\}} \M{P}^{(i)} (\M{H}_k^ {\{i\}})^\top +\M{H}_k^ {\{l\}}  \M{P}^{(l)} (\M{H}_k^ {\{l\}})^\top$ for each pair $i,l$.
\end{description}

In both Assumptions \ref{ass_Pu_constA} and \ref{ass_Pu_constB}, the coupling probability at time $k$ between two objects $i$ and $l$ is constant with respect to the state values of the respective objects. Based on either assumption \ref{ass_Pu_constA} or \ref{ass_Pu_constB}, analogous definitions follow for $\tQuk^{il} = 1-\tPuk^{il}$, $\tVuk{\Set{O}} = \prod_{1\leq i< l \leq \no} \tQuk^{il}$ and $\truk{\Set{A}}$. Employing these assumptions and dropping the time index $k$ in \eqref{eq_full_post3} leads to

\begin{align}
   f( \munderbar{\V{x}}_{}, \V{a}, \munderbar{\V{b}} \vert \munderbar{\V{z}}_{})  \propto &  \ist\ist \Psi \big(\V{a}\rmv\rmv, \munderbar{\V{b}} \big) \Bigg[ \prod_{i=1}^{\no} f_0^{(i)}\big(\V{x}^{(i)}\big) \Bigg] \nn\\
   &\hspace{0mm} \times  q^{(0)}\big(  \Set{S}(\V{b}^{(0)})  \big) \Bigg[ \prod_{j=1}^{m} q^{(j)}\big( \munderbar{\V{x}} , \Set{S}(\V{b}^{(j)}) \big)  \Bigg]  \label{eq_full_post4}
\end{align}
where, for any subset $\Set{I}\subseteq \Set{O}$, the $\{q^{(j)}\}_{j=0}^m$ factors become
\begin{equation*}
 q^{(0)}\big( \Set{I}  \big)  =  \sum_{\Set{P} \in \mathfrak{B}({{\Set{I}})}}  \prod_{\Set{U} \in \Set{P} } \big( 1- \Pd{\Set{U}} \big) \,  \tru{\Set{U}}  
\end{equation*}
and for $j\geq 1$
\begin{align}
 q^{(j)}\big(\munderbar{\V{x}} , \Set{I}\big)  =  \begin{cases}
  \frac{ \tru{\Set{I}}  \, \Pd{\Set{I}} \, \fu{\Set{I}} (\V{z}^{(j)} \vert \munderbar{\V{x}}^{\Set{I}} ) }{\lfa \ffa (\V{z}^{(j)})}, & \text{if } \Set{I} \neq \emptyset \\
    1, & \text{if } \Set{I} = \emptyset. \nonumber 
\end{cases}
\end{align}


The specialization of the LBP algorithm to the Gaussian case leads to the Gaussian LBP (GLBP) message passing algorithm. Using standard Gaussian results \cite[Ch. 3.8]{RisAruGor:B04}, the GLBP measurement evaluation for $\V{z}^{(j)}$ and a vector $\V{b}$ such that $\Set{S}(\V{b}) = \Set{I} \neq \emptyset$ becomes 
\begin{align}
    \varphi_{\munderbar{\V{x}} \V{b}^{j}}(\V{b})   = \frac{ \tru{\Set{I}}   \, \Pd{\Set{I}} \, \fg \big( \V{z}^{(j)}; \sum_{i \in \Set{I}} \M{H}^{\Set{I}} \V{m}^{(i)}, \M{S}_{\Set{I}} \big) }{\lfa \ffa (\V{z}^{(j)}) }\label{eq_gauss_phib}
\end{align}
where $\M{S}_{\Set{I}}  =  \M{R}^{{\Set{I}}} + \sum_{i\in \Set{I}} \M{H}^{{\Set{I}}}   \M{P}^{(i)}\big[ \M{H}^{{\Set{I}}} \big]^\top$, whereas $\varphi_{\munderbar{\V{x}} \V{b}^{0}}(\V{b}) = q^{(0)}(\Set{S}(\V{b}))$ for all $\V{b}$. The data association loop in the GLBP is identical to the general LBP of the previous section. 

The marginal likelihood of object $i$, when group $\Set{I} \supseteq \{i\}$ is associated with the $j$-th measurement becomes the weighted Gaussian likelihood function 
\begin{IEEEeqnarray}{rCl}
   \xi_i^{(j)}(\V{x}^{(i)}, \Set{I})  = \frac{ \tru{\Set{I}}   \, \Pd{\Set{I}} \, \fg \big( \V{z}^{(j)}_{\Set{I}\setminus i};  \M{H}^{{\Set{I}}} \V{x}^{(i)}, \M{S}_{\Set{I}\setminus i}  \big) }{\lfa \ffa (\V{z}^{(j)}) }  
\end{IEEEeqnarray}
where 
\begin{IEEEeqnarray*}{rCl}
\V{z}^{(j)}_{\Set{I}\setminus i} &=& \V{z}^{(j)} - \sum_{l\in \Set{I}\setminus i} \M{H}^{{\Set{I}}} \V{m}^{(l)} \\
 \M{S}_{\Set{I}\setminus i} &=& \M{S}_{\Set{I}}   - \M{H}^{{\Set{I}}}   \M{P}^{(i)}\big[ \M{H}^{{\Set{I}}} \big]^\top \,.
\end{IEEEeqnarray*}
From the above marginal-likelihood and \eqref{eq_gamma_qj}, each message $\gamma_{{q}^j  \V{x}^{i} }(\cdot)$ is a mixture of $2^{\no-1}$ weighted Gaussian likelihood components and one constant component. Thus, the belief at each node $i$ from \eqref{eq_belif_x} is also a mixture distribution with an exponential growth in the number of components, \ie, $O(2^{\no m})$. To avoid this increase in the number of components, after each update with a $\gamma_{{q}^j  \V{x}^{i} }(\cdot)$ message in the belief \eqref{eq_belif_x}, we perform a projection from the space of Gaussian mixture distributions to that of single Gaussian distributions. For any object $i$, recall that $ {f}_{0}^{(i)}\big(\V{x}^{(i)} \big)$ is the predicted density $  {f}\big(\V{x}_k^{(i)} \big\vert \munderbar{\V{z}}_{1:k-1}\big)$ and which, in virtue of \ref{ass_dyn_g}, has a single Gaussian form. We sequentially perform for $j=1, 2, \cdots, m$: 
 \begin{equation}
\begin{cases}
\tilde{f}_{j}^{(i)} (\V{x}^{(i)})   & \propto f_{j-1}^{(i)} (\V{x}^{(i)})\,  \gamma_{{q}^{j}  \V{x}^{i} }(\V{x}^{(i)}),\\
f_{j}^{(i)} (\V{x}^{(i)})   & = \big( \Pi \: \tilde{f}_{j}^{(i)}    \big) (\V{x}^{(i)})
\label{eq_gauss_up_proj}
\end{cases}
\end{equation}
\noindent where $\tilde{f}_{j}^{(i)} (\V{x}^{(i)}) $ has a Gaussian mixture form $\fgm (\V{x}^{(i)}; \munderbar{\V{m}}_j^{(i)}, \munderbar{\M{P}}_j^{(i)})$ with $\munderbar{\V{m}}_j^{(i)}$ and $\munderbar{\M{P}}_j^{(i)}$ given by standard Gaussian results \cite[Ch. 3.8]{RisAruGor:B04}. Furthermore, $\Pi$ is the projection defined in \eqref{eq_gm_proj}. At the last step ($j=m$), we obtain the single Gaussian approximation $f_{m}^{(i)} (\V{x}^{(i)})$ to the marginal distribution at node $\V{x}^{(i)}$. Note that while the message $ q^{(0)}( \cdot  )$ is necessary for the data association loop, the message $\gamma_{{q}^0  \V{x}^{i} }(\V{x}^{(i)})$ is a constant and has no influence on the update of \eqref{eq_gauss_up_proj}. 

\subsection{Complexity Analysis}

The measurement evaluation step of GLBP algorithm incurs a computational complexity on the order of $O(m\, 2^{\no})$, following from the computation of \eqref{eq_gauss_phib} for all $j$ and all $\V{b}$. Here, a computational complexity $O(2^{\no ^2})$, stemming from evaluating $\tru{\Set{I}} $ for all sets $\Set{I}\subset \Set{O}$, may be ignored since evaluating either \eqref{eq_def_tPuA} or \eqref{eq_def_tPuB} may be achieved much faster than \eqref{eq_gauss_phib}. The computational complexity of the data association loop remains at $O(m\,\no\, 2^{\no})$. The computational complexity of the update in \eqref{eq_gauss_up_proj} is $O(m\, \no\, 2^{\no})$ for all objects. This follows, since for each object $i$ and each measurement index $j$, a Gaussian mixture with $2^{\no-1}$ terms is formed after an update with a message $ \gamma_{{q}^{j}  \V{x}^{i} }$ is performed, followed by a projection to a single Gaussian which also involves $O(2^{\no})$ computations. 


\section{Numerical Results} \label{sec:results}

In this section, we present numerical results that compare our proposed GLBP algorithm with a Gaussian Exact-Marginals (GEM) algorithm. The two algorithms share all elements, such as, Gaussian object models, model parameters, approximations (i.e., \ref{ass_dyn_g}-\ref{ass_obs_g} and \ref{ass_Pu_constB}), with the exception that GLBP employs our proposed LBP mechanism to evaluate the marginal association probabilities ${p}_i({a})$ $\forall {a}$ whereas the GEM performs an exact evaluation of these marginals by a direct evaluation of the joint probabilities of all multiobject association events. The GEM algorithm may be seen as a variant of the method of~\cite{SveUlmHam:12} for unresolved measurements, the difference being that the latter method avoids multiobject association events that are deemed impossible for a given sensor in order to reduce its computational complexity. The GLBP and GEM methods are compared in a static and a dynamic scenario, as presented next.  
  
\subsection{Static Scenario}

\begin{figure}[t!]
\centering
\vspace{-0.5mm}
\psfrag{Simulation setup}[l][l][1]{\raisebox{0mm}{\hspace{0mm}}}
\psfrag{Unresolved Measurementttt}[l][l][0.88]{\raisebox{-0mm}{\hspace{0mm}Unresolved measurement}}
\psfrag{Resolved Measurement}[l][l][0.88]{\raisebox{-0mm}{\hspace{0mm}Resolved measurement}}
\psfrag{Clutter Measurement}[l][l][0.88]{\raisebox{-0mm}{\hspace{0mm}Clutter measurement}}
\psfrag{Targets}[l][l][0.88]{\raisebox{-0mm}{\hspace{0mm}Object position}}
\psfrag{graph12345678901}[l][l][0.88]{\raisebox{-0mm}{\hspace{-0.1mm}Resolution graph}}

\psfrag{X coordinate [m]}[l][l][1]{\raisebox{-6mm}{\hspace{0mm}X coordinate [m]}}
\psfrag{Y coordinate [m]}[l][l][1]{\raisebox{4mm}{\hspace{0mm}Y coordinate [m]}}

\psfrag{30}[l][l][0.95]{\raisebox{0mm}{\hspace{0mm}$30$}}
\psfrag{20}[l][l][0.95]{\raisebox{0mm}{\hspace{0mm}$20$}}
\psfrag{10}[l][l][0.95]{\raisebox{0mm}{\hspace{0mm}$10$}}
\psfrag{0}[l][l][0.95]{\raisebox{0mm}{\hspace{0mm}$0$}}
\psfrag{-30}[l][l][0.95]{\raisebox{0mm}{\hspace{-2mm}$-30$}}
\psfrag{-20}[l][l][0.95]{\raisebox{0mm}{\hspace{-2mm}$-20$}}
\psfrag{-10}[l][l][0.95]{\raisebox{0mm}{\hspace{-2mm}$-10$}}

\psfrag{0.002}[l][l][0.95]{\raisebox{0mm}{\hspace{-0mm}$0.002$}}
\psfrag{0.004}[l][l][0.95]{\raisebox{0mm}{\hspace{-0mm}$0.004$}}
\psfrag{0.006}[l][l][0.95]{\raisebox{0mm}{\hspace{-0mm}$0.006$}}
\psfrag{0.008}[l][l][0.95]{\raisebox{0mm}{\hspace{-0mm}$0.008$}}
\psfrag{0.01}[l][l][0.95]{\raisebox{0mm}{\hspace{-0mm}$0.01$}}
\psfrag{0.012}[l][l][0.95]{\raisebox{0mm}{\hspace{-0mm}$0.012$}}
\psfrag{0.014}[l][l][0.95]{\raisebox{0mm}{\hspace{-0mm}$0.014$}}
\psfrag{0.016}[l][l][0.95]{\raisebox{0mm}{\hspace{-0mm}$0.016$}}
\psfrag{0.018}[l][l][0.95]{\raisebox{0mm}{\hspace{-0mm}$0.018$}}
\psfrag{0.02}[l][l][0.95]{\raisebox{0mm}{\hspace{-0mm}$0.02$}}

 \includegraphics[scale=0.30,draft=false,clip=true]{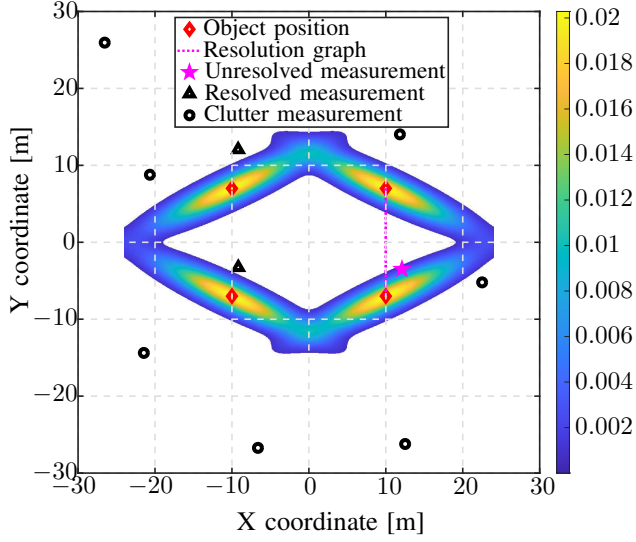}
\caption{Static scenario depicting four closely-spaced objects with their location probability densities, which are indicated by the colored areas. Also, note an instance of the resolution graph and of the different measurements. The graph and the type of measurements are unknown to the filter.}
\label{fig_setup1}
\vspace{0mm}
\end{figure}
A static (or fixed time) scenario is considered where the object state vectors correspond to the XY coordinates of the objects and are Gaussian distributed as displayed in Figure~\ref{fig_setup1}. A linear Gaussian observation model \eqref{eq_gobs_model} is employed to generate noisy location measurements from the objects through the process of object grouping followed by group detection and measurement generation (described in Sec.~\ref{sec:model}). Further model parameters are: $\Au = 10^2\, \M{I}_2$; $\Pdk{\Set{I}} = 0.9$ $\forall \Set{I}$; clutter observations are uniformly distributed on the compact space $[-30,\, 30] \times [-30,\, 30]$; for an object group $\Set{I}$, the observation matrix is taken to be $ \M{H}_k^{\Set{I}} = \card{\Set{I}}^{-1} \, \M{I}_2$ which calculates the centroid average of the object positions; whereas the observation noise covariance matrix exhibits a slow increase with the cardinality $\card{\Set{I}}$ of the object group (here, taken to be $ \M{R}_k^{\Set{I}} = \sigma^2 \sqrt[3]{\card{\Set{I}}}\, \M{I}_2$). The data-association loop (step 2 of the generic LBP algorithm) is run until $\max_{ij}  \big\vert \log \frac{\Mabt{\ell+1}_{ij}}{\Mabt{\ell}_{ij}} \big\vert \leq 10^{-9}$, for some $\ell$, or otherwise until the maximum number of iterations $L=50$ is reached.

In order to quantify the performance of association probability estimation, the~\ac{tvd} is used to compare the GLBP marginal probabilities $\{ \hat{p}_i(\cdot)\}_{i=1}^{\no}$ and the exact probabilities $\{  {p}_i(\cdot)\}_{i=1}^{\no}$ obtained using GEM. The~\ac{tvd}~\cite[Ch.\,5]{DevLug:B01} is a distance between two probability mass functions (pmfs). Specifically, for two pmfs $\hat{p}$ and $p$ defined on the same discrete space of outcomes $\Set{\Omega}$, their~\ac{tvd} $\dtv (\hat{p},p)$ is given by $\dtv (\hat{p},p) = \frac{1}{2} \sum_{\omega \in \Set{\Omega}} \vert \hat{p}(\omega)- p(\omega) \vert$, with $\dtv ( \hat{p},p) \in [0,\,1]$. When the~\ac{tvd} between two pmfs is equal to a specific value, say $v$, then the difference in probability between the two pmfs is upper bounded by $v$ for any outcome $\omega \in \Omega$.

For each object $i$, the GLBP algorithm provides the association pmf estimate $\hat{p}_i$ given in \eqref{eq_pa_est}, defined on the space of outcomes $\Omega = \{ 0,1,2, \cdots, m_k\}$ and where $\hat{p}_i(j)$ is the estimated posterior probability that object $i$ contributed to measurement $j$, if $j>0$, and the estimated posterior probability that object $i$ is misdetected if $j=0$. Since both GLBP and GEM algorithms provide an association pmf estimate for each object, we employ an \ac{atvd} metric between their respective pmf sets. Specifically letting $\{ \hat{p}_i(\cdot)\}_{i=1}^{\no}$ and $\{  {p}_i(\cdot)\}_{i=1}^{\no}$ be the pmfs produced by GLBP and GEM, the \ac{atvd} is defined as $\datv (\hat{p},  {p}) = \frac{1}{\no} \sum_{i=1}^{\no} \dtv (\hat{p}_i, {p}_i)$.  

\begin{figure}[t!]
\centering
\vspace{-0.5mm}
\psfrag{BP Distance from BruteF Marginal PMF}[l][l][1]{\raisebox{0mm}{\hspace{0mm}}}
\psfrag{Noise Variance}[c][c][1]{\raisebox{-6mm}{\hspace{-0mm}Resolved-measurement Noise Variance $\sigma^2$ [m$^2$]}}
\psfrag{Total Variation Distance}[c][c][1]{\raisebox{4mm}{\hspace{2mm}Average \ac{tvd}}}
\psfrag{Median}[l][l][0.9]{\raisebox{-0mm}{\hspace{0.0mm}Median}}
\psfrag{Percentiles}[l][l][0.9]{\raisebox{-0mm}{\hspace{0.0mm}Percentiles}}

\psfrag{30}[l][l][0.95]{\raisebox{-3mm}{\hspace{0mm}$30$}}
\psfrag{20}[l][l][0.95]{\raisebox{-3mm}{\hspace{0mm}$20$}}
\psfrag{10}[l][l][0.95]{\raisebox{-3mm}{\hspace{0mm}$10$}}
\psfrag{40}[l][l][0.95]{\raisebox{-3mm}{\hspace{0mm}$40$}}
\psfrag{50}[l][l][0.95]{\raisebox{-3mm}{\hspace{-0mm}$50$}}
\psfrag{60}[l][l][0.95]{\raisebox{-3mm}{\hspace{-0mm}$60$}}
\psfrag{70}[l][l][0.95]{\raisebox{-3mm}{\hspace{-0mm}$70$}}
\psfrag{80}[l][l][0.95]{\raisebox{-3mm}{\hspace{-0mm}$80$}}
\psfrag{90}[l][l][0.95]{\raisebox{-3mm}{\hspace{-0mm}$90$}}
\psfrag{100}[l][l][0.95]{\raisebox{-3mm}{\hspace{-0mm}$100$}}

\psfrag{0.005}[l][l][0.95]{\raisebox{0mm}{\hspace{-1mm} }}
\psfrag{0.01}[l][l][0.95]{\raisebox{0mm}{\hspace{-2mm}$0.01$}}
\psfrag{0.015}[l][l][0.95]{\raisebox{0mm}{\hspace{-1mm} }}
\psfrag{0.02}[l][l][0.95]{\raisebox{0mm}{\hspace{-2mm}$0.02$}}
\psfrag{0.025}[l][l][0.95]{\raisebox{0mm}{\hspace{-1mm} }}
\psfrag{0.03}[l][l][0.95]{\raisebox{0mm}{\hspace{-2mm}$0.03$}}
\psfrag{0.04}[l][l][0.95]{\raisebox{0mm}{\hspace{-2mm}$0.04$}}
\psfrag{0.035}[l][l][0.95]{\raisebox{0mm}{\hspace{-1mm}}}
\psfrag{0.08}[l][l][0.95]{\raisebox{0mm}{\hspace{-2mm}$0.08$}}
\psfrag{0.1}[l][l][0.95]{\raisebox{0mm}{\hspace{-2mm}$0.1$}}
\psfrag{0.12}[l][l][0.95]{\raisebox{0mm}{\hspace{-2mm}$0.12$}}

 \includegraphics[scale=0.31,draft=false,clip=true]{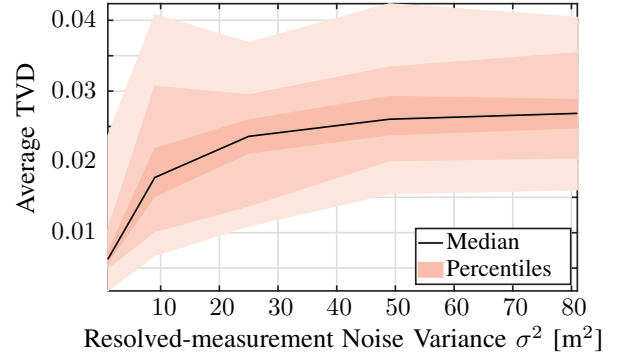}
\caption{Static scenario depicting average TVD between the GLBP estimated marginal probabilities and the exact evaluation of the marginals. The curves correspond to the median error and $25\%-75\%$ percentile bands. The clutter rate is $\lfak = 10$. }
\label{fig_tvd_noise_std}
\vspace{0mm}
\end{figure}

\begin{figure}[t!]
\centering
\vspace{-0.5mm}
\psfrag{BP Distance from BruteF Marginal PMF}[l][l][1]{\raisebox{0mm}{\hspace{0mm}}}
\psfrag{Noise Variance}[c][c][1]{\raisebox{-6mm}{\hspace{-0mm}Clutter rate $\lfak$}}
\psfrag{Total Variation Distance}[c][c][1]{\raisebox{6mm}{\hspace{2mm}Average \ac{tvd}}}
\psfrag{Median}[l][l][0.9]{\raisebox{-0mm}{\hspace{0.0mm}Median}}
\psfrag{Percentiles}[l][l][0.9]{\raisebox{-0mm}{\hspace{0.0mm}Percentiles}}

\psfrag{30}[l][l][0.95]{\raisebox{-3mm}{\hspace{0mm}$30$}}
\psfrag{25}[l][l][0.95]{\raisebox{-3mm}{\hspace{0mm}$25$}}
\psfrag{20}[l][l][0.95]{\raisebox{-3mm}{\hspace{0mm}$20$}}
\psfrag{15}[l][l][0.95]{\raisebox{-3mm}{\hspace{0mm}$15$}}
\psfrag{10}[l][l][0.95]{\raisebox{-3mm}{\hspace{0mm}$10$}}
\psfrag{5}[l][l][0.95]{\raisebox{-3mm}{\hspace{0mm}$5$}}
\psfrag{40}[l][l][0.95]{\raisebox{0mm}{\hspace{0mm}$40$}}
\psfrag{50}[l][l][0.95]{\raisebox{0mm}{\hspace{-0mm}$50$}}
\psfrag{60}[l][l][0.95]{\raisebox{0mm}{\hspace{-0mm}$60$}}
\psfrag{70}[l][l][0.95]{\raisebox{0mm}{\hspace{-0mm}$70$}}
\psfrag{80}[l][l][0.95]{\raisebox{0mm}{\hspace{-0mm}$80$}}
\psfrag{90}[l][l][0.95]{\raisebox{0mm}{\hspace{-0mm}$90$}}
\psfrag{100}[l][l][0.95]{\raisebox{0mm}{\hspace{-0mm}$100$}}

\psfrag{0.01}[l][l][0.95]{\raisebox{0mm}{\hspace{-2mm}$0.01$}}
\psfrag{0.02}[l][l][0.95]{\raisebox{0mm}{\hspace{-2mm}$0.02$}}
\psfrag{0.03}[l][l][0.95]{\raisebox{0mm}{\hspace{-2mm}$0.03$}}
\psfrag{0.04}[l][l][0.95]{\raisebox{0mm}{\hspace{-2mm}$0.04$}}
\psfrag{0.05}[l][l][0.95]{\raisebox{0mm}{\hspace{-2mm}$0.05$}}
\psfrag{0.06}[l][l][0.95]{\raisebox{0mm}{\hspace{-2mm}$0.06$}}
\psfrag{0.07}[l][l][0.95]{\raisebox{0mm}{\hspace{-2mm}$0.07$}}
\psfrag{0.08}[l][l][0.95]{\raisebox{0mm}{\hspace{-2mm}$0.08$}}
\psfrag{0.1}[l][l][0.95]{\raisebox{0mm}{\hspace{-2mm}$0.1$}}
\psfrag{0.12}[l][l][0.95]{\raisebox{0mm}{\hspace{-2mm}$0.12$}}

 \includegraphics[scale=0.31,draft=false,clip=true]{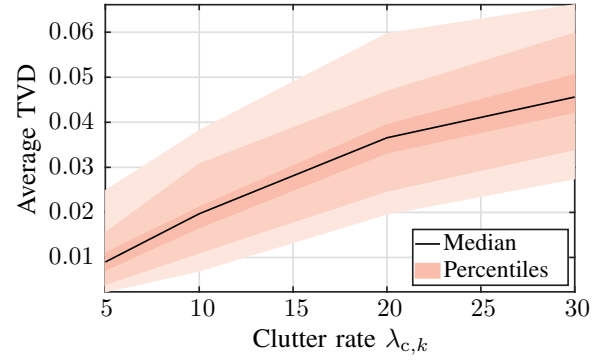}
\caption{Static scenario depicting average TVD between the GLBP estimated marginals and the exact evaluation of the marginals. The curves correspond to the median error and $25\%-75\%$ percentile bands. The measurement noise has $\sigma^2 = 10$. }
\label{fig_tdd_noise_clutter}
\vspace{-3mm}
\end{figure}

Performance curves depicting the \ac{atvd} between the GLBP estimated association probabilities and the GEM exact association probabilities are depicted in Fig.~\ref{fig_tvd_noise_std} and Fig.~\ref{fig_tdd_noise_clutter}, as functions of the measurement noise parameter $\sigma^2$ and clutter rate $\lfak$, respectively. Both figures showcase the median \ac{atvd} and various percentile intervals obtained from a number of $300$ independent simulations. Note that the GLBP estimated association pmfs are close in \ac{atvd} to the GEM exact association pmfs and that this distance slowly increases with the measurement noise variance and clutter rate.

The lower computational complexity of the GLBP algorithm, as compared to the GEM algorithm, is reflected in the high speed-ups reported in Table~\ref{tab_static}, which are defined as the ratios between the wall-clock run-times of the GEM and the GLBP algorithm. The small loss in ~\ac{atvd}  is compensated by this high increase in computational efficiency of the GLBP. The computational speed-up becomes more significant at increased clutter rates or number of objects, while increasing the latter leads to the most pronounced increase. This is in line with the results in Sec.~\ref{sec_comp_complexity} since the GEM algorithm incurs a computational complexity of $O(m^{\no})$ whereas the GLBP has a computational complexity of $O(m\, \no 2^{\no})$.

\begin{table}[h!]
\begin{tabular}{l|c|cccc}
\cline{2-6}
\rowcolor[HTML]{FFFFFF} 
\cellcolor[HTML]{FFFFFF}                   & \cellcolor[HTML]{FFFFFF}                                                                               & \multicolumn{4}{c}{\cellcolor[HTML]{FFFFFF}Clutter Rate}                                                                                                                                                                                    \\ \cline{3-6} 
\rowcolor[HTML]{EFEFEF} 
\multirow{-2}{*}{\cellcolor[HTML]{FFFFFF}} & \multirow{-2}{*}{\cellcolor[HTML]{FFFFFF}\begin{tabular}[c]{@{}c@{}}Number of\\  objects\end{tabular}} & \multicolumn{1}{c|}{\cellcolor[HTML]{EFEFEF}$\lambda=5$} & \multicolumn{1}{c|}{\cellcolor[HTML]{EFEFEF}$\lambda=10$} & \multicolumn{1}{l|}{\cellcolor[HTML]{EFEFEF}$\lambda=20$} & \multicolumn{1}{l}{\cellcolor[HTML]{EFEFEF}$\lambda=30$} \\ \hline
\rowcolor[HTML]{FFFFFF} 
\ac{atvd}                                      & \cellcolor[HTML]{EFEFEF}                                                                               & \multicolumn{1}{c|}{\cellcolor[HTML]{FFFFFF}0.01}        & \multicolumn{1}{c|}{\cellcolor[HTML]{FFFFFF}0.02}         & \multicolumn{1}{c|}{\cellcolor[HTML]{FFFFFF}0.04}         & 0.05                                                      \\
\rowcolor[HTML]{ECF4FF} 
Speed-up                                   & \multirow{-2}{*}{\cellcolor[HTML]{EFEFEF}$\no=4$}                                                      & \multicolumn{1}{c|}{\cellcolor[HTML]{ECF4FF}7}           & \multicolumn{1}{c|}{\cellcolor[HTML]{ECF4FF}31}           & \multicolumn{1}{c|}{\cellcolor[HTML]{ECF4FF}212}           & 746                                                       \\ \hline
\rowcolor[HTML]{FFFFFF} 
\ac{atvd}                                       & \cellcolor[HTML]{EFEFEF}                                                                               & \multicolumn{1}{c|}{\cellcolor[HTML]{FFFFFF}0.04}          & \multicolumn{1}{c|}{\cellcolor[HTML]{FFFFFF}0.06}           & \multicolumn{1}{c|}{\cellcolor[HTML]{FFFFFF}0.1}             &  \multicolumn{1}{c}{\cellcolor[HTML]{FFFFFF}0.14}                                                         \\
\rowcolor[HTML]{ECF4FF} 
Speed-up                                   & \multirow{-2}{*}{\cellcolor[HTML]{EFEFEF}$\no=5$}                                                      & \multicolumn{1}{c|}{\cellcolor[HTML]{ECF4FF}13}          & \multicolumn{1}{c|}{\cellcolor[HTML]{ECF4FF}91}           & \multicolumn{1}{c|}{\cellcolor[HTML]{ECF4FF}1234}             &      \multicolumn{1}{c}{\cellcolor[HTML]{ECF4FF}6730}                                                      \\ \hline
\end{tabular}
\vspace{1mm}
\caption{\ac{atvd} between the proposed~GLBP and GEM algorithms and speed-up values (ratio of GEM and GLBP wall-clock run times). The resolved noise variance is kept at $\sigma^2 =4$ m$^2$.}
\label{tab_static}
\vspace{-5mm}
\end{table}

\newcommand{\Ts}{T_{\mathrm{s}}}

\subsection{Dynamic Scenario}

\begin{figure}[t!]
\hspace{-5mm}
\centering
\vspace{-0.7mm}
\psfrag{X coordinate [m]}[l][l][1]{\raisebox{-6mm}{\hspace{0mm}X coordinate [m]}}
\psfrag{Y coordinate [m]}[c][c][1]{\raisebox{4mm}{\hspace{-0mm}Y coordinate [m]}}
\psfrag{Target tracks}[l][l][0.8]{\raisebox{0mm}{\hspace{0mm}True object tracks}}
\psfrag{Clutter measurement}[l][l][0.8]{\raisebox{-0mm}{\hspace{0.0mm}Clutter measurements}}
\psfrag{Rezolved measurement}[l][l][0.8]{\raisebox{-0mm}{\hspace{0.0mm}Resolved measurements}}
\psfrag{Unrezolved measurement}[l][l][0.8]{\raisebox{-0mm}{\hspace{0.0mm}Unresolved measurements}}

\psfrag{50}[l][l][0.95]{\raisebox{0mm}{\hspace{-0mm}$50$}}
\psfrag{-50}[l][l][0.95]{\raisebox{0mm}{\hspace{-2mm}$-50$}}
\psfrag{100}[l][l][0.95]{\raisebox{0mm}{\hspace{-0mm}$100$}}
\psfrag{-100}[l][l][0.95]{\raisebox{0mm}{\hspace{-2mm}$-100$}}
\psfrag{-150}[l][l][0.95]{\raisebox{0mm}{\hspace{-2mm}$-150$}}
\psfrag{150}[l][l][0.95]{\raisebox{0mm}{\hspace{-0mm}$150$}}

\psfrag{0}[l][l][0.95]{\raisebox{0mm}{\hspace{-1mm}$0$}}
\psfrag{0.02}[l][l][0.95]{\raisebox{0mm}{\hspace{-1mm}$0.02$}}
\psfrag{0.03}[l][l][0.95]{\raisebox{0mm}{\hspace{-1mm}$0.03$}}
\psfrag{0.04}[l][l][0.95]{\raisebox{0mm}{\hspace{-1mm}$0.04$}}
\psfrag{0.05}[l][l][0.95]{\raisebox{0mm}{\hspace{-1mm}$0.05$}}
\psfrag{0.06}[l][l][0.95]{\raisebox{0mm}{\hspace{-1mm}$0.06$}}
\psfrag{0.07}[l][l][0.95]{\raisebox{0mm}{\hspace{-1mm}$0.07$}}
\psfrag{0.08}[l][l][0.95]{\raisebox{0mm}{\hspace{-1mm}$0.08$}}
\psfrag{0.1}[l][l][0.95]{\raisebox{0mm}{\hspace{-1mm}$0.1$}}
\psfrag{0.12}[l][l][0.95]{\raisebox{0mm}{\hspace{-1mm}$0.12$}}

 \includegraphics[scale=0.40,draft=false,clip=true]{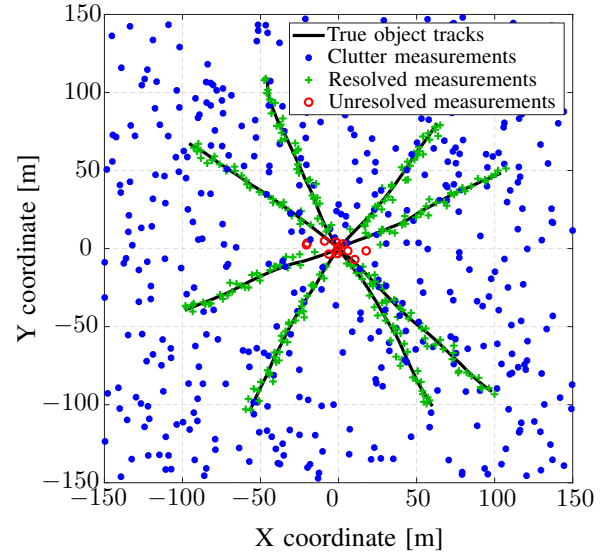}
\caption{Dynamic scenario depicting an instance of four object tracks. The object tracks are designed so as to meet in the center $(0,0)$ of the 2D plane. Also shown are the measurements and measurement types---clutter, resolved (generated by a single object), and unresolved (generated by two or more objects). The measurement noise has parameter $\sigma^2 = 5$ and $\lfak = 5$. Note that the measurement type information is not available to the proposed filter.}
\label{fig_dyn_truth}
\vspace{0mm}
\end{figure}

\begin{figure}[b!]
\hspace{-8mm}
\vspace{-5.5mm}
\centering%
\psfrag{X coordinate [m]}[l][l][1]{\raisebox{-6mm}{\hspace{0mm}X coordinate [m]}}
\psfrag{Y coordinate [m]}[c][c][1]{\raisebox{4mm}{\hspace{-0mm}Y coordinate [m]}}
\psfrag{Truth}[l][l][0.83]{\raisebox{0mm}{\hspace{0mm}True}}
\psfrag{Est1}[l][l][0.83]{\raisebox{-0mm}{\hspace{0.0mm}Est. $1$}}
\psfrag{Est2}[l][l][0.83]{\raisebox{-0mm}{\hspace{0.0mm}Est. $2$}}
v\psfrag{Est3}[l][l][0.83]{\raisebox{-0mm}{\hspace{0.0mm}Est. $3$}}
\psfrag{Est4}[l][l][0.83]{\raisebox{-0mm}{\hspace{0.0mm}Est. $4$}}
\psfrag{Est5}[l][l][0.83]{\raisebox{-0mm}{\hspace{0.0mm}Est. $5$}}%

\psfrag{50}[l][l][0.95]{\raisebox{0mm}{\hspace{-0mm}$50$}}
\psfrag{-50}[l][l][0.95]{\raisebox{0mm}{\hspace{-1mm}$-50$}}
\psfrag{100}[l][l][0.95]{\raisebox{0mm}{\hspace{-0mm}$100$}}
\psfrag{-100}[l][l][0.95]{\raisebox{0mm}{\hspace{-1mm}$-100$}}
\psfrag{-150}[l][l][0.95]{\raisebox{0mm}{\hspace{-1mm}$-150$}}
\psfrag{150}[l][l][0.95]{\raisebox{0mm}{\hspace{-0mm}$150$}}%
\psfrag{0}[l][l][0.95]{\raisebox{0mm}{\hspace{-1mm}$0$}}
\psfrag{0.02}[l][l][0.95]{\raisebox{0mm}{\hspace{-1mm}$0.02$}}
\psfrag{0.03}[l][l][0.95]{\raisebox{0mm}{\hspace{-1mm}$0.03$}}
\psfrag{0.04}[l][l][0.95]{\raisebox{0mm}{\hspace{-1mm}$0.04$}}
\psfrag{0.05}[l][l][0.95]{\raisebox{0mm}{\hspace{-1mm}$0.05$}}
\psfrag{0.06}[l][l][0.95]{\raisebox{0mm}{\hspace{-1mm}$0.06$}}
\psfrag{0.07}[l][l][0.95]{\raisebox{0mm}{\hspace{-1mm}$0.07$}}
\psfrag{0.08}[l][l][0.95]{\raisebox{0mm}{\hspace{-1mm}$0.08$}}
\psfrag{0.1}[l][l][0.95]{\raisebox{0mm}{\hspace{-1mm}$0.1$}}
\psfrag{0.12}[l][l][0.95]{\raisebox{0mm}{\hspace{-1mm}$0.12$}}%
\includegraphics[scale=0.36,draft=false,clip=true]{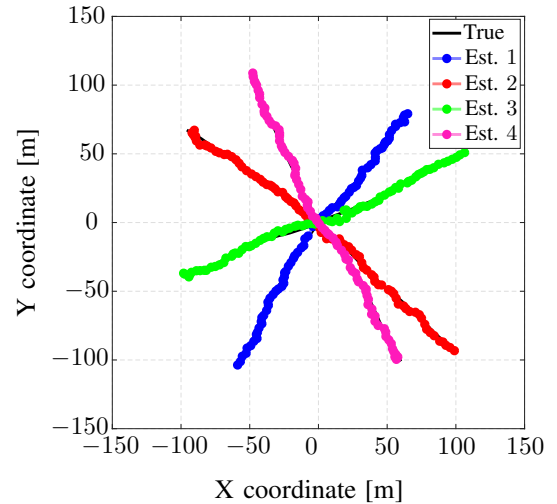}
\caption{GLBP estimated object tracks corresponding to the case in Fig. \ref{fig_dyn_truth}. Ground truth tracks are labeled as ``true'' and the four estimated tracks are labeled as ``Est.'' $1$ through $4$.}
\label{fig_dyn_est}
\vspace{0mm}
\end{figure}

In this section, we consider a scenario in which the state vector of any object $i\in \Set{O}$ evolves according to the discrete white-noise acceleration model~\cite[Ch. 6.3.2]{BarRonKir:B01} in a 2D plane. More specifically, the individual object state vectors contain object 2D position $(x,y)$ and velocity information $(\dot{x},\dot{y})$, \ie, $\V{x}^{(i)} = \big[ x, y, \dot{x}, \dot{y} \big]^\top$. Considering a sampling period $\Ts$, state transitions are characterized by the Gaussian transition kernel $f^{(i)}_{k+1\vert k}(\V{x} \vert \V{y}) = \fg (\V{x}; \M{F}\,\V{y}, \M{Q}) $, where 
\begin{equation*}
\M{F} = 
\begin{bmatrix}
	\M{I}_2  & \Ts\, \M{I}_2           \\
	\M{0}_2  & \M{I}_2    
\end{bmatrix}
\quad \text{and} \quad \M{Q} = \sigma_{o}^2
\begin{bmatrix}
	\frac{\Ts^4}{4}\M{I}_2  & \frac{\Ts^3}{2} \M{I}_2           \\
	\frac{\Ts^3}{2}\M{I}_2  & \Ts^2 \M{I}_2   
\end{bmatrix}	      \,. 
\end{equation*}
The specific filtering parameters employed in this work are $\Ts =1$s and an object noise of $ \sigma_{o} = 5\times10^{-3}$ m/s$^2$. For an object group $\Set{I}$, the observation model of \eqref{eq_gobs_model} has observation matrix $ \M{H}_k^{\Set{I}} = \card{\Set{I}}^{-1} \, \big[ \M{I}_2, \M{0}_2\big]$ which yields the centroid average of the object positions; whereas the observation noise covariance matrix is $ \M{R}_k^{\Set{I}} = \sigma^2 \M{I}_2$ if ${\Set{I}}$ is a singleton and $ \M{R}_k^{\Set{I}} = 2\sigma^2 \M{I}_2$ whenever $\card{\Set{I}}>1$. The probability of detection is a constant $\Pdk{\Set{I}} = 0.98$ for any group $  \Set{I}\subseteq \Set{O}$ of objects. Clutter observations are uniformly distributed on the compact $[-150,\, 150] \times [-150,\, 150]$. The sensor resolution capabilities are described by the matrix $\Au = 10^2\, \M{I}_2$. An instance of four object tracks with the unresolved measurement model is depicted in Fig.~\ref{fig_dyn_truth}. Correspondingly, the estimated object tracks produced by the GLBP filter are showcased in Fig.~\ref{fig_dyn_est}.

\begin{figure}[t!]
\hspace{-2mm}
\vspace{-5mm}
\centering
\psfrag{X coordinate [m]}[l][l][1]{\raisebox{-6mm}{\hspace{0mm}X coordinate [m]}}
\psfrag{Y coordinate [m]}[c][c][1]{\raisebox{4mm}{\hspace{-0mm}Y coordinate [m]}}
\psfrag{Truth}[l][l][0.83]{\raisebox{0mm}{\hspace{0mm}True}}
\psfrag{Est1}[l][l][0.83]{\raisebox{-0mm}{\hspace{0.0mm}Est. $1$}}
\psfrag{Est2}[l][l][0.83]{\raisebox{-0mm}{\hspace{0.0mm}Est. $2$}}
v\psfrag{Est3}[l][l][0.83]{\raisebox{-0mm}{\hspace{0.0mm}Est. $3$}}
\psfrag{Est4}[l][l][0.83]{\raisebox{-0mm}{\hspace{0.0mm}Est. $4$}}
\psfrag{Est5}[l][l][0.83]{\raisebox{-0mm}{\hspace{0.0mm}Est. $5$}}

\psfrag{50}[l][l][0.95]{\raisebox{0mm}{\hspace{-0mm}$50$}}
\psfrag{-50}[l][l][0.95]{\raisebox{0mm}{\hspace{-2mm}$-50$}}
\psfrag{100}[l][l][0.95]{\raisebox{0mm}{\hspace{-0mm}$100$}}
\psfrag{-100}[l][l][0.95]{\raisebox{0mm}{\hspace{-2mm}$-100$}}
\psfrag{-150}[l][l][0.95]{\raisebox{0mm}{\hspace{-2mm}$-150$}}
\psfrag{150}[l][l][0.95]{\raisebox{0mm}{\hspace{-0mm}$150$}}
\psfrag{0}[l][l][0.95]{\raisebox{0mm}{\hspace{-1mm}$0$}}
\psfrag{0.02}[l][l][0.95]{\raisebox{0mm}{\hspace{-1mm}$0.02$}}
\psfrag{0.03}[l][l][0.95]{\raisebox{0mm}{\hspace{-1mm}$0.03$}}
\psfrag{0.04}[l][l][0.95]{\raisebox{0mm}{\hspace{-1mm}$0.04$}}
\psfrag{0.05}[l][l][0.95]{\raisebox{0mm}{\hspace{-1mm}$0.05$}}
\psfrag{0.06}[l][l][0.95]{\raisebox{0mm}{\hspace{-1mm}$0.06$}}
\psfrag{0.07}[l][l][0.95]{\raisebox{0mm}{\hspace{-1mm}$0.07$}}
\psfrag{0.08}[l][l][0.95]{\raisebox{0mm}{\hspace{-1mm}$0.08$}}
\psfrag{0.1}[l][l][0.95]{\raisebox{0mm}{\hspace{-1mm}$0.1$}}
\psfrag{0.12}[l][l][0.95]{\raisebox{0mm}{\hspace{-1mm}$0.12$}}
\includegraphics[scale=0.36,draft=false,clip=true]{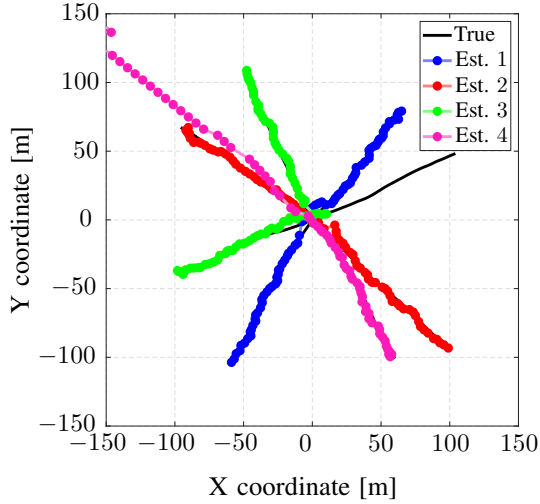}
\caption{Estimated object tracks using the No-U GLBP algorithm that assumes at most one-to-one associations and does not employ any model for unresolved measurements. One may observe the divergence of track Est. $4$ and the switch of track Est. $3$ to following a different object when the objects become closely spaced.}
\label{fig_dyn_est_nogum}
\vspace{-3mm}
\end{figure}

In Fig.~\ref{fig_dyn_est_nogum}, we showcase the estimated tracks produced by a Gaussian LBP algorithm that does not employ any unresolved measurement model, but instead assumes the standard one-to-one association rule. We name this method as the No-U GLBP algorithm. As can be seen from Fig.~\ref{fig_dyn_est_nogum}, the No-U GLBP algorithm struggles with handling closely-spaced objects, leading to track divergence and track switching. 

\begin{figure}[b!]
\hspace{-5mm}
\vspace{-3mm}
\centering
\psfrag{Median Oracle1234567}[l][l][0.8]{\raisebox{0mm}{\hspace{0mm}Median Oracle}}
\psfrag{P-interval Oracle}[l][l][0.8]{\raisebox{-0mm}{\hspace{-0mm}P-int. Oracle}}
\psfrag{Median brute}[l][l][0.8]{\raisebox{0mm}{\hspace{-0.0mm}Median GEM-JPDAF}}
\psfrag{P-interval brute}[l][l][0.8]{\raisebox{-0mm}{\hspace{0.0mm}P-int. GEM-JPDAF}}
\psfrag{Median BP}[l][l][0.8]{\raisebox{-0mm}{\hspace{0.0mm}Median GLBP}}
\psfrag{P-interval BP}[l][l][0.8]{\raisebox{-0mm}{\hspace{0.0mm}P-int. GLBP}}

\psfrag{30}[l][l][0.95]{\raisebox{-1mm}{\hspace{0mm}$30$}}
\psfrag{25}[l][l][0.95]{\raisebox{-1mm}{\hspace{0mm}$25$}}
\psfrag{20}[l][l][0.95]{\raisebox{-1mm}{\hspace{0mm}$20$}}
\psfrag{5}[l][l][0.95]{\raisebox{-1mm}{\hspace{0mm}$5$}}
\psfrag{15}[l][l][0.95]{\raisebox{-1mm}{\hspace{0mm}$15$}}
\psfrag{10}[l][l][0.95]{\raisebox{-1mm}{\hspace{-0.3mm}$10$}}
\psfrag{40}[l][l][0.95]{\raisebox{-1mm}{\hspace{0mm}$40$}}
\psfrag{50}[l][l][0.95]{\raisebox{-1mm}{\hspace{-0mm}$50$}}
\psfrag{60}[l][l][0.95]{\raisebox{-1mm}{\hspace{-0mm}$60$}}
\psfrag{70}[l][l][0.95]{\raisebox{-1mm}{\hspace{-0mm}$70$}}
\psfrag{80}[l][l][0.95]{\raisebox{-1mm}{\hspace{-0mm}$80$}}
\psfrag{90}[l][l][0.95]{\raisebox{-1mm}{\hspace{-0mm}$90$}}
\psfrag{100}[l][l][0.95]{\raisebox{-1mm}{\hspace{-0mm}$100$}}

\psfrag{time}[l][l][0.9]{\raisebox{-0mm}{\hspace{0.0mm} Time [s]}}
\psfrag{MSE}[l][l][0.9]{\raisebox{1mm}{\hspace{-10.0mm} Labeled MSE [m$^2$]}}

\includegraphics[scale=0.28,draft=false,clip=true]{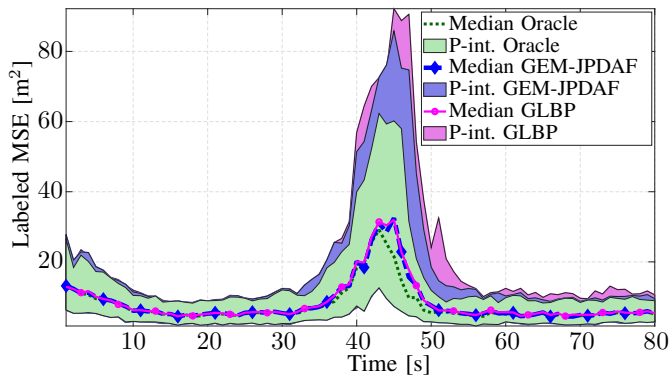}
\caption{Labeled MSE of estimated object tracks (2D position only) as a function of time. The curves correspond to the median error and $10\%-90\%$ percentile (P-int.) bands. The measurement noise has $\sigma^2 = 10$ and the clutter rate is $\lfak = 5$.}
\label{fig_dyn_mse}
\vspace{0mm}
\end{figure}
We compare the performance of our proposed method GLBP with three reference methods---No-U GLBP, GEM-JPDAF, and the Oracle filter. The GEM-JPDAF employs the exact data-association step of GEM in a \ac{jpdaf} framework. Thus, GEM-JPDAF is similar to the method of~\cite{SveUlmHam:12} with two modifications. Namely, the GEM-JPDAF does not employ any strategies to eliminate association hypotheses in the calculation of marginal probabilities and approximates object marginal pdfs as Gaussian. The Oracle filter employs the true associations between objects and measurements followed by a marginalization step in order to update each object posterior distribution, which is also assumed Gaussian.   

In Fig.~\ref{fig_dyn_mse}, the Labeled Mean Squared Error (LMSE) between the estimated object tracks and the true object tracks are shown as a function of time for the three methods. The LMSE is obtained as the sum of individual-object MSE curves, as a function of time, and divided by the number of objects. Individual-object MSE curves are calculated between each true object track and the corresponding estimated track (i.e., having the same label as the true object track). The median and percentile curves in Fig.~\ref{fig_dyn_mse} are computed from a set of $100$ independent runs. One can observe that all LMSE curves have sharp peaks when the objects meet in the middle of the 2D plane. Notice that the LMSE values of the GLBP are close to those of the GEM-JPDAF and slightly worse than those of the Oracle filter, only noticeable when the object tracks cross and the association problem is at its hardest to solve. 

\begin{table}[h!]
\begin{tabular}{ccccccc}
\cline{3-7}
\rowcolor[HTML]{FFFFFF} 
\multicolumn{1}{l}{\cellcolor[HTML]{FFFFFF}}                               & \cellcolor[HTML]{FFFFFF}                          & \multicolumn{5}{c}{\cellcolor[HTML]{FFFFFF}Observation noise variance $\sigma^2$ [m$^2$]}                                                                                                                                                                                                                                                \\ \cline{3-7} 
\rowcolor[HTML]{EFEFEF} 
\multicolumn{1}{l}{\multirow{-2}{*}{\cellcolor[HTML]{FFFFFF}}}             & \multirow{-2}{*}{\cellcolor[HTML]{FFFFFF}}        & \multicolumn{1}{m{0.035\textwidth}|}{\cellcolor[HTML]{EFEFEF} \centering{$2$}} & \multicolumn{1}{m{0.035\textwidth}|}{\cellcolor[HTML]{EFEFEF}\centering{$5$}} & \multicolumn{1}{m{0.04\textwidth}|}{\cellcolor[HTML]{EFEFEF}\centering{$10$}} & \multicolumn{1}{m{0.035\textwidth}|}{\cellcolor[HTML]{EFEFEF}\centering{$20$}} & \multicolumn{1}{m{0.035\textwidth}}{\cellcolor[HTML]{EFEFEF}\centering{$40$}} \\ \hline
\rowcolor[HTML]{FFFFFF} 
\multicolumn{1}{c|}{\cellcolor[HTML]{FFFFFF}}                              & \multicolumn{1}{c|}{\cellcolor[HTML]{FFFFFF}ALMSE} & \multicolumn{1}{c|}{\cellcolor[HTML]{FFFFFF}2.8}             & \multicolumn{1}{c|}{\cellcolor[HTML]{FFFFFF}5.4}           & \multicolumn{1}{c|}{\cellcolor[HTML]{FFFFFF}9.6}           & \multicolumn{1}{c|}{\cellcolor[HTML]{FFFFFF}15.6}            & 26                                                       \\
\rowcolor[HTML]{ECF4FF} 
\multicolumn{1}{c|}{\multirow{-2}{*}{\cellcolor[HTML]{FFFFFF}GLBP}}    & \multicolumn{1}{c|}{\cellcolor[HTML]{ECF4FF}IQR}  & \multicolumn{1}{c|}{\cellcolor[HTML]{ECF4FF}1.4}             & \multicolumn{1}{c|}{\cellcolor[HTML]{ECF4FF}1.8}            & \multicolumn{1}{c|}{\cellcolor[HTML]{ECF4FF}4.7}            & \multicolumn{1}{c|}{\cellcolor[HTML]{ECF4FF}6}             &11.2                                                         \\ \hline%
\rowcolor[HTML]{FFFFFF} 
\multicolumn{1}{c|}{\cellcolor[HTML]{FFFFFF}}                              & \multicolumn{1}{c|}{\cellcolor[HTML]{FFFFFF}ALMSE} & \multicolumn{1}{c|}{\cellcolor[HTML]{FFFFFF}7.2}             & \multicolumn{1}{c|}{\cellcolor[HTML]{FFFFFF}8.8}           & \multicolumn{1}{c|}{\cellcolor[HTML]{FFFFFF}11.2}           & \multicolumn{1}{c|}{\cellcolor[HTML]{FFFFFF}18.2}            & 36.2                                                       \\
\rowcolor[HTML]{ECF4FF} 
\multicolumn{1}{c|}{\multirow{-2}{*}{\cellcolor[HTML]{FFFFFF}No-U GLBP}}    & \multicolumn{1}{c|}{\cellcolor[HTML]{ECF4FF}IQR}  & \multicolumn{1}{c|}{\cellcolor[HTML]{ECF4FF}197}             & \multicolumn{1}{c|}{\cellcolor[HTML]{ECF4FF}186}            & \multicolumn{1}{c|}{\cellcolor[HTML]{ECF4FF}90}            & \multicolumn{1}{c|}{\cellcolor[HTML]{ECF4FF}189}             & 545                                                         \\ \hline%
\rowcolor[HTML]{FFFFFF} 
\multicolumn{1}{c|}{\cellcolor[HTML]{FFFFFF}}                              & \multicolumn{1}{c|}{\cellcolor[HTML]{FFFFFF}ALMSE} & \multicolumn{1}{c|}{\cellcolor[HTML]{FFFFFF}2.9}             & \multicolumn{1}{c|}{\cellcolor[HTML]{FFFFFF}5.4}           & \multicolumn{1}{c|}{\cellcolor[HTML]{FFFFFF}9.4}           & \multicolumn{1}{c|}{\cellcolor[HTML]{FFFFFF}15.6}            & 25.3                                                        \\
\rowcolor[HTML]{ECF4FF} 
\multicolumn{1}{c|}{\multirow{-2}{*}{\cellcolor[HTML]{FFFFFF}GEM-JPDAF}} & \multicolumn{1}{c|}{\cellcolor[HTML]{ECF4FF}IQR}  & \multicolumn{1}{c|}{\cellcolor[HTML]{ECF4FF}1.4}             & \multicolumn{1}{c|}{\cellcolor[HTML]{ECF4FF}2.1}            & \multicolumn{1}{c|}{\cellcolor[HTML]{ECF4FF}4.3}            & \multicolumn{1}{c|}{\cellcolor[HTML]{ECF4FF}5.9}             & 12.3                                                         \\ \hline
\rowcolor[HTML]{FFFFFF} 
\multicolumn{1}{c|}{\cellcolor[HTML]{FFFFFF}}                              & \multicolumn{1}{c|}{\cellcolor[HTML]{FFFFFF}ALMSE} & \multicolumn{1}{c|}{\cellcolor[HTML]{FFFFFF}2.6}             & \multicolumn{1}{c|}{\cellcolor[HTML]{FFFFFF}4.9}            & \multicolumn{1}{c|}{\cellcolor[HTML]{FFFFFF}8.3}           & \multicolumn{1}{c|}{\cellcolor[HTML]{FFFFFF}13.4}            & 22                                                        \\
\rowcolor[HTML]{ECF4FF} 
\multicolumn{1}{c|}{\multirow{-2}{*}{\cellcolor[HTML]{FFFFFF}Oracle Filter}}     & \multicolumn{1}{c|}{\cellcolor[HTML]{ECF4FF}IQR}  & \multicolumn{1}{c|}{\cellcolor[HTML]{ECF4FF}1.3}             & \multicolumn{1}{c|}{\cellcolor[HTML]{ECF4FF}1.6}            & \multicolumn{1}{c|}{\cellcolor[HTML]{ECF4FF}2.9}            & \multicolumn{1}{c|}{\cellcolor[HTML]{ECF4FF}3.2}             & 5.2                                                         \\ \hline
\end{tabular}
\vspace{1mm}
\caption{Median and IQR values for the time-averaged LMSE (ALMSE) of various methods as a function of different observation noise variance. The clutter rate is fixed at $\lfak = 5$.}
\label{tab_var}
\vspace{-3mm}
\end{table}

In Tables \ref{tab_var} and \ref{tab_clutter}, we present the LMSE as function of observation noise variance and clutter rate, respectively. The curve of LMSE as a function of time is compressed into a single time-average LMSE value, denoted ALMSE, with the median and interquartile range (IQR) being reported in the tables. Additionally, the speed-up of the proposed GLBP with respect to the GEM-JPDAF is reported in Table \ref{tab_clutter}. From both Tables \ref{tab_var} and \ref{tab_clutter}, one may observe that the GLBP performance is similar to that of the GEM-JPDAF, but at a fraction of the computational run time. The high IQR values of the No-U GLBP method are due to track divergences. 

\begin{table}[h!]
\begin{tabular}{cccccc}
\cline{3-6}
\rowcolor[HTML]{FFFFFF} 
\multicolumn{1}{c}{\cellcolor[HTML]{FFFFFF}}                               & \cellcolor[HTML]{FFFFFF}                   & \multicolumn{4}{c}{\cellcolor[HTML]{FFFFFF}Clutter Rate}                                                                                                                                                                                    \\ \cline{3-6} 
\rowcolor[HTML]{EFEFEF} 
\multicolumn{1}{l}{\multirow{-2}{*}{\cellcolor[HTML]{FFFFFF}}}             & \multirow{-2}{*}{\cellcolor[HTML]{FFFFFF}} & \multicolumn{1}{c|}{\cellcolor[HTML]{EFEFEF}$\lambda=10$} & \multicolumn{1}{c|}{\cellcolor[HTML]{EFEFEF}$\lambda=20$} & \multicolumn{1}{l|}{\cellcolor[HTML]{EFEFEF}$\lambda=30$} & \multicolumn{1}{l}{\cellcolor[HTML]{EFEFEF}$\lambda=40$} \\ \hline
\rowcolor[HTML]{FFFFFF} 
\multicolumn{1}{c|}{\cellcolor[HTML]{FFFFFF}}                              & ALMSE                                       & \multicolumn{1}{|c|}{\cellcolor[HTML]{FFFFFF}2.7}         & \multicolumn{1}{c|}{\cellcolor[HTML]{FFFFFF}2.8}          & \multicolumn{1}{c|}{\cellcolor[HTML]{FFFFFF}3.2}          & 3.3                                                      \\
\rowcolor[HTML]{ECF4FF} 
\multicolumn{1}{c|}{\cellcolor[HTML]{FFFFFF}}                              & IQR                                        & \multicolumn{1}{|c|}{\cellcolor[HTML]{ECF4FF}1.2}         & \multicolumn{1}{c|}{\cellcolor[HTML]{ECF4FF}1.2}          & \multicolumn{1}{c|}{\cellcolor[HTML]{ECF4FF}1.5}          & 2                                                       \\
\rowcolor[HTML]{FFCCC9} 
\multicolumn{1}{c|}{\multirow{-3}{*}{\cellcolor[HTML]{FFFFFF}GLBP}}    & Speed-Up                                   & \multicolumn{1}{|c|}{\cellcolor[HTML]{FFCCC9}213}          & \multicolumn{1}{c|}{\cellcolor[HTML]{FFCCC9}1131}          & \multicolumn{1}{c|}{\cellcolor[HTML]{FFCCC9}3355}         & 7973                                                      \\ \hline
\rowcolor[HTML]{FFFFFF}
\multicolumn{1}{c|}{\cellcolor[HTML]{FFFFFF}}                              & ALMSE                                       & \multicolumn{1}{|c|}{\cellcolor[HTML]{FFFFFF}6.9}         & \multicolumn{1}{c|}{\cellcolor[HTML]{FFFFFF}9.1}          & \multicolumn{1}{c|}{\cellcolor[HTML]{FFFFFF}127}          & 390                                                      \\
\rowcolor[HTML]{ECF4FF} 
\multicolumn{1}{c|}{\multirow{-2}{*}{\cellcolor[HTML]{FFFFFF}No-U GLBP}} & IQR                                        & \multicolumn{1}{|c|}{\cellcolor[HTML]{ECF4FF}426}         & \multicolumn{1}{c|}{\cellcolor[HTML]{ECF4FF}799}          & \multicolumn{1}{c|}{\cellcolor[HTML]{ECF4FF}1265}          & 1650                                                      \\ \hline
\rowcolor[HTML]{FFFFFF}  
\multicolumn{1}{c|}{\cellcolor[HTML]{FFFFFF}}                              & ALMSE                                       & \multicolumn{1}{|c|}{\cellcolor[HTML]{FFFFFF}2.7}         & \multicolumn{1}{c|}{\cellcolor[HTML]{FFFFFF}2.8}          & \multicolumn{1}{c|}{\cellcolor[HTML]{FFFFFF}3.2}          & 3.2                                                       \\
\rowcolor[HTML]{ECF4FF} 
\multicolumn{1}{c|}{\multirow{-2}{*}{\cellcolor[HTML]{FFFFFF}GEM-JPDAF}} & IQR                                        & \multicolumn{1}{|c|}{\cellcolor[HTML]{ECF4FF}1.2}         & \multicolumn{1}{c|}{\cellcolor[HTML]{ECF4FF}1.3}          & \multicolumn{1}{c|}{\cellcolor[HTML]{ECF4FF}1.6}          & 1.6                                                       \\ \hline
\rowcolor[HTML]{FFFFFF} 
\multicolumn{1}{c|}{\cellcolor[HTML]{FFFFFF}}                              & ALMSE                                       & \multicolumn{1}{|c|}{\cellcolor[HTML]{FFFFFF}2.5}         & \multicolumn{1}{c|}{\cellcolor[HTML]{FFFFFF}2.5}          & \multicolumn{1}{c|}{\cellcolor[HTML]{FFFFFF}2.6}          & 2.6                                                       \\
\rowcolor[HTML]{ECF4FF} 
\multicolumn{1}{c|}{\multirow{-2}{*}{\cellcolor[HTML]{FFFFFF}Oracle Filter}}     & IQR                                        & \multicolumn{1}{|c|}{\cellcolor[HTML]{ECF4FF}0.8}         & \multicolumn{1}{c|}{\cellcolor[HTML]{ECF4FF}0.8}          & \multicolumn{1}{c|}{\cellcolor[HTML]{ECF4FF}0.9}          & 0.7                                                       \\ \hline
\end{tabular}
\vspace{1mm}
\caption{Median and IQR values for the time-averaged LMSE (ALMSE) of various methods as a function of different clutter rates. The observation noise variance is fixed at $\sigma^2 = 2$.}
\label{tab_clutter}
\vspace{-3mm}
\end{table}

A measure for track switches may be constructed as follows. Apart from the labeled MSE between estimated and ground truth object states, the best matching assignment---in terms of total MSE---between the unlabeled estimated set of states and the ground truth set of object states may be evaluated for any method at any time step. We refer to this MSE value as the best-matching MSE. Such linear assignment problems are routinely solved using the Hungarian algorithm \cite{PapSte:B98}. In this work, a track switch is declared whenever a difference between the labeled MSE and the best-matching MSE is greater in magnitude than the respective best-matching MSE. In Table \ref{tab_switches}, we present the average number of such track switches that occur throughout an $80$-second simulation, again averaged over $100$ independent simulations. Note the similar values for the proposed GLBP and GEM-JPDAF methods, whereas the No-U GLBP method exhibits higher number of track switches that corroborate the higher ALMSE values from Table \ref{tab_var}.

\begin{table}[t!]
\begin{tabular}{cccccc}
\cline{2-6}
\rowcolor[HTML]{FFFFFF} 
\multicolumn{1}{l}{\cellcolor[HTML]{FFFFFF}}                   & \multicolumn{5}{c}{\cellcolor[HTML]{FFFFFF}Observation Noise Variance}                                                                                                                                                                                                                                              \\ \cline{2-6} 
\rowcolor[HTML]{EFEFEF} 
\multicolumn{1}{l}{\multirow{-2}{*}{\cellcolor[HTML]{FFFFFF}}} & \multicolumn{1}{c|}{\cellcolor[HTML]{EFEFEF}$\sigma^2 = 2$} & \multicolumn{1}{c|}{\cellcolor[HTML]{EFEFEF}$\sigma^2 =5$} & \multicolumn{1}{l|}{\cellcolor[HTML]{EFEFEF}$\sigma^2 =10$} & \multicolumn{1}{l|}{\cellcolor[HTML]{EFEFEF}$\sigma^2 = 20$} & \multicolumn{1}{l}{\cellcolor[HTML]{EFEFEF}$\sigma^2 = 40$} \\ \hline
\rowcolor[HTML]{FFFFFF} 
\multicolumn{1}{c|}{\cellcolor[HTML]{FFFFFF}GLBP}              & \multicolumn{1}{c|}{\cellcolor[HTML]{FFFFFF}0.5}            & \multicolumn{1}{c|}{\cellcolor[HTML]{FFFFFF}1.1}           & \multicolumn{1}{c|}{\cellcolor[HTML]{FFFFFF}1.2}            & \multicolumn{1}{c|}{\cellcolor[HTML]{FFFFFF}1.6}             & 2                                                           \\
\rowcolor[HTML]{ECF4FF} 
\multicolumn{1}{c|}{\cellcolor[HTML]{ECF4FF}No-U GLBP}         & \multicolumn{1}{c|}{\cellcolor[HTML]{ECF4FF}4.9}            & \multicolumn{1}{c|}{\cellcolor[HTML]{ECF4FF}3.5}           & \multicolumn{1}{c|}{\cellcolor[HTML]{ECF4FF}5.4}            & \multicolumn{1}{c|}{\cellcolor[HTML]{ECF4FF}7.9}             & 7.1                                                         \\
\rowcolor[HTML]{FFFFFF} 
\multicolumn{1}{c|}{\cellcolor[HTML]{FFFFFF}GEM-JPDAF}             & \multicolumn{1}{c|}{\cellcolor[HTML]{FFFFFF}0.5}            & \multicolumn{1}{c|}{\cellcolor[HTML]{FFFFFF}1.2}           & \multicolumn{1}{c|}{\cellcolor[HTML]{FFFFFF}1.3}            & \multicolumn{1}{c|}{\cellcolor[HTML]{FFFFFF}1.7}             & 1.2                                                         \\ \hline
\end{tabular}
\vspace{1mm}
\caption{Average number of track switches encountered by each method over the course of a single simulation. The clutter rate is fixed at $\lfak = 5$.}
\label{tab_switches}
\vspace{-3mm}
\end{table}

\section{Conclusion}

In this paper, we address the data association problem and object state inference from potentially unresolved measurements, where two or more objects may jointly generate a single measurement. Building upon a well-known unresolved measurement model, a probability distribution for partitions of objects is developed, which serves as a prior distribution for the data-association problem. Furthermore, we propose a message-passing algorithm for performing state inference with the resulting model. Through numerical experiments, we demonstrate the effectiveness of the proposed method and its reduced computational cost as compared to a direct marginalization method. 

\section*{Acknowledgment}

We thank Prof. Simon Maskell for the constructive talks on the subject of unresolved measurements.   

\appendices

\section{Proof of Proposition \ref{prop_ruI}} \label{app_prop_ruI}
    Without loss of generality, let $\Set{P}$ be a partition $\Set{P} =\{\Set{U}_1, \Set{U}_2, \cdots, \Set{U}_m\} \in \mathfrak{B}({\Set{O}})$ (with arbitrary $1\leq m \leq \no$). Then, the summation in \eqref{eq_prob_part_def} over graphs with connected components identical to the sets in $\Set{P}$ leads to  
 \begin{IEEEeqnarray}{rCl}
\mathbb{P}_{\RS{P}}& & \big( \{\Set{U}_1, \Set{U}_2, \cdots, \Set{U}_m\} \big\vert \munderbar{\V{x}}_k \big) \nonumber \\
 &&= \Vuk{\Set{O}}(\munderbar{\V{x}}_k)  \sum_{ \substack{ G\in \mathfrak{G}({\Set{O}}) \\ \Set{C}({G}) = \Set{P}}}  \prod_{(i,l) \in \mathcal{E}_{{G}}} \frac{\Puk(\V{x}_k^{(i)}, \V{x}_k^{(l)})}{\Quk(\V{x}_k^{(i)}, \V{x}_k^{(l)})} \nonumber \\
&&= \Vuk{\Set{O}}(\munderbar{\V{x}}_k)  \prod_{j=1}^m \bigg[ \sum_{ \substack{  G_j \in \mathfrak{G}({\Set{U}_j}) \\ \Set{C}({G_j}) = \Set{U}_j }}   \prod_{(i,l) \in \mathcal{E}_{G_j} } \frac{\Puk(\V{x}_k^{(i)}, \V{x}_k^{(l)})}{\Quk(\V{x}_k^{(i)}, \V{x}_k^{(l)})} \bigg] \nonumber \\
&&= \Vuk{\Set{O}}(\munderbar{\V{x}}_k)  \prod_{j=1}^m \big[ \ruk{\Set{U}_j} ( \munderbar{\V{x}}_k^{\Set{U}_j} ) \big] \,. \nonumber
 \end{IEEEeqnarray}     
The second equality above follows from the observation that a graph $G\in \mathfrak{G}({\Set{O}})$ that satisfies $ \Set{C}({G}) = \Set{P}$  may be seen as composed of the independent subgraphs $\{G_j\}_{j=1}^m$, each being connected and where no path exists between two subgraphs $G_j$ and $G_{j'}$ with $j\neq j'$. Thus, the set of edges $\mathcal{E}_{{G}}$ may be partitioned into $\mathcal{E}_{{G}} = \mathcal{E}_{{G}_1} \cup  \mathcal{E}_{{G}_2} \cup \cdots \cup  \mathcal{E}_{{G}_m}$ according to these subgraphs. Finally, the summation over graphs $G$ becomes a summation over independent subgraphs $G_j$ with $ \Set{C}({G_j}) = \Set{U}_j$ for each $j\in \intset{1}{m}$.    

\section{Proof of Proposition~\ref{prop_prior1_aB}}
\label{app_prior_a}
Assume throughout this section that $\munderbar{\V{b}}_k$ is a valid association tuple, i.e., there exists an $\V{a}_k \in \Set{A}_k$ such that $\Psi_k(\V{a}_k, \munderbar{\V{b}}_k) = 1$. In contrast to classical multiobject tracking (where only one object can contribute to a single measurement), here, several objects may jointly generate a measurement. Indeed, the set $\Set{S}(\munderbar{\V{b}}_k) $ contains all detected groups of objects. Moreover, the elements of $\Set{S}(\munderbar{\V{b}}_k) $ may be viewed as \emph{virtual objects} or \emph{clusters}, to which the classical assignment rule of one distinct measurement per virtual object applies. The association variable $\munderbar{\V{b}}_k$ also conveys the knowledge of the set of detected objects $ \Set{A}^{\mathrm d}_{ } = \cup_{j} \, \Set{S}(\V{b}_k^{(j)})$ and that of the set of misdetected objects $ \Set{A}^{\mathrm m}_{ } = \Set{S}(\V{b}_k^{(0)})$. Note that $\munderbar{\V{b}}_k$ does not contain any information on the partitioning of the misdetected object set. In fact, the partitioning of $ \Set{A}^{\mathrm m}_{ } $ remains a hidden variable in the proposed model.


Consider an arbitrary set of detected object groups $\Set{A} $ with $\card{\Set{A}} \leq m_k$, \ie, there exists a partition $\Set{P}\in \mathfrak{B}({\Set{O}})$ such that $\Set{A} \subseteq \Set{P}_{}$. Note that for such a set $\Set{A}$, there are potentially multiple corresponding assignment tuples $\munderbar{\V{b}}_k$, namely, all valid $\munderbar{\V{b}}_k \in {\Set{B}}_k$ with $\Set{S}(\munderbar{\V{b}}_k) = \Set{A}$. Indeed, in this case the groups in $\Set{A}$ can be seen as the virtual objects. Mirroring the reasoning in~\cite[p. 315]{BarLi:B95}, it follows that the number of valid assignment tuples $\munderbar{\V{b}}_k$ for which the same set of virtual objects is detected is given by the number of permutations of the $m_k$ measurements taken as $\card{\Set{A}}$. This corresponds to the number of events under which a distinct measurement index from the set $\intset{1}{m_k}$ is assigned to an element from $\Set{A}$---a virtual object. Assuming that these permutations are a priori equally likely, for a valid  $\munderbar{\V{b}}_k \in {\Set{B}}_k$ and $\card{\Set{A}}\leq m_k$, one has
\begin{IEEEeqnarray}{rCl}
    p(\munderbar{\V{b}}_k \vert \Set{A}, m_k) &=& \bigg[ \frac{m_k !}{ (m_k- \card{\Set{A}} )! } \bigg]^{-1} \delta_{\Set{A}} (\Set{S}(\munderbar{\V{b}}_k))\nonumber \\
    &=& \bigg[ \frac{m_k !}{ m^{\fa}_{k}! }\bigg]^{-1}\delta_{\Set{A}} (\Set{S}(\munderbar{\V{b}}_k))
    \label{eq_prob_misI}
\end{IEEEeqnarray}
where the number of false alarms is $m^{\fa}_{k} = m_k - \card{\Set{S}(\munderbar{\V{b}}_k)}$.


Let $\Set{P}\in \mathfrak{B}({\Set{O}})$ be any partition of objects, denote the indices of detected objects via the set $\Set{A}^{\mathrm d} $ and the set of indices of misdetected objects via $\Set{A}^{\mathrm m} = \Set{O} \setminus \Set{A}^{\mathrm d}$. From Assumption \ref{ass_unres_mb} and given $\Set{P}$, the probability of detecting the objects with indices in $\Set{A}^{\mathrm d} $ and misdetecting those with indices in $\Set{A}^{\mathrm m} $ is  
\begin{IEEEeqnarray}{rCll}
\vspace{-.3cm}p(\Set{A}^{\mathrm d}, \Set{A}^{\mathrm m} \vert \Set{P}, \munderbar{\V{x}}_k) &=& \prod_{\Set{U}\in \Set{P}} \Big[ &\indicator{\Set{A}^{\mathrm d}}(\Set{U})  \Pdk{\ist \Set{U}}(\munderbar{\V{x}}_k^{\Set{U}})  \nn \\ 
&& & +  \indicator{\Set{A}^{\mathrm m}}(\Set{U}) \big(1- \Pdk{\ist \Set{U}}(\munderbar{\V{x}}_k^{\Set{U}})\big) \Big] . \label{eq_pd_groups}
\end{IEEEeqnarray}
Note that the above probability is zero whenever the hypothesis in Assumption \ref{ass_unres_mb} is not respected, that is, whenever any element $\Set{U}$ of $\Set{P}$ contains both detected and misdetected it holds that $\indicator{\Set{A}^{\mathrm d}}(\Set{U}) = \indicator{\Set{A}^{\mathrm m}}(\Set{U}) =0$.   

For any $n\leq m_k$, let $\Set{A} =\{\Set{A}_1, \Set{A}_2, \cdots, \Set{A}_{n} \}$ be a set of detected object groups, i.e., 
$\Set{A}_i \subset \Set{O}$ $\forall i$, $\Set{A}_i\cap \Set{A}_j = \emptyset$ for any $i\neq j$, and all the objects in $\Set{A}^{\mathrm d}\triangleq  \cup_i \Set{A}_i$ are detected while the objects in $\Set{A}^{\mathrm m} = \Set{O} \setminus \Set{A}^{\mathrm d}$ are misdetected. Alternatively, the set $\Set{A}$ may be seen as a partial partition of $\Set{O}$ (\ie, there exists a partition $\Set{P}\in \mathfrak{B}({\Set{O}})$ such that $\Set{A} \subseteq \Set{P}_{}$) that contains all the detected object groups.  Furthermore, marginalizing over the unknown partitions of the misdetected objects, the prior probability of $\Set{A}$ can be expressed as
\begin{IEEEeqnarray}{rCl}
    && \hspace{-0.3cm} p(\Set{A} \vert \munderbar{\V{x}}_k)  \nonumber \\
   &=&  \sum_{\Set{U} \in \mathfrak{B}({{\Set{A}}^{\mathrm m}})} p(\Set{A},\, \Set{U} \vert \munderbar{\V{x}}_k) \nonumber \\
    & = & \sum_{\Set{U} \in \mathfrak{B}({{\Set{A}}^{\mathrm m}})} p( \Set{A} \cup \Set{U}, \Set{A}^{\mathrm d}, \Set{A}^{\mathrm m} \vert \munderbar{\V{x}}_k) \nonumber \\
    & = & \sum_{\Set{U} \in \mathfrak{B}({{\Set{A}}^{\mathrm m}})} p( \Set{A}^{\mathrm d}, \Set{A}^{\mathrm m} \vert \Set{A} \cup \Set{U} ,\munderbar{\V{x}}_k) \mathbb{P}_{\RS{P}}(\Set{A} \cup \Set{U}\vert  \munderbar{\V{x}}_k) \nonumber \\
 & = & \sum_{\Set{U} \in \mathfrak{B}({{\Set{A}}^{\mathrm m}})}  \Bigg[  \prod_{i=1}^{n} \Pdk{\ist \Set{A}_i}(\munderbar{\V{x}}_k^{\ist \Set{A}_i}) \Bigg]  \Bigg[  \prod_{\Set{W} \in \Set{U}}\big(1-\Pdk{\ist \Set{W}}(\munderbar{\V{x}}_k^{\ist \Set{W}})\big) \Bigg]    \nonumber \\
    &  &  \hspace{1cm} \times  \Vuk{\Set{O}}(\munderbar{\V{x}}_k) \Bigg[  \prod_{i=1}^{n}  \ruk{\Set{A}_i} (\munderbar{\V{x}}_k^{\Set{A}_i} ) \Bigg]  \prod_{\Set{W}\in \Set{U}}  \ruk{\Set{W}} (\munderbar{\V{x}}_k^{\Set{W}} )  \nonumber \\
    & = & \Vuk{\Set{O}}(\munderbar{\V{x}}_k) \Bigg[  \prod_{i=1}^{n} \Pdk{\ist \Set{A}_i}(\munderbar{\V{x}}_k^{\ist \Set{A}_i})\,  \ruk{\Set{A}_i} (\munderbar{\V{x}}_k^{\Set{A}_i} ) \Bigg] \nonumber \\ 
    && \hspace{1cm} \times \sum_{\Set{U} \in \mathfrak{B}({{\Set{A}}^{\mathrm m}})}  \prod_{\Set{W}\in \Set{U}} \big( 1- \Pdk{\ist \Set{W}}(\munderbar{\V{x}}_k^{\Set{W}})\big) \,  \ruk{\Set{W}} (\munderbar{\V{x}}_k^{\Set{W}} )  \label{eq_p_det_groups} 
\end{IEEEeqnarray}
where in the second line we employed the fact that knowing $\Set{A}$ and $\Set{U}$ is equivalent to knowing both: (i) the partition $ \Set{A} \cup \Set{U}$ of all the objects in $\Set{O}$; and (ii) the sets of detected $\Set{A}^{\mathrm d}$ and misdetected $\Set{A}^{\mathrm m}$ objects. The penultimate line follows from Proposition \ref{prop_ruI} and \eqref{eq_pd_groups}, by plugging in the probability of a partition given the object states $\munderbar{\V{x}}_k$ and the probabilities of detection/misdetection for object groups. 

Note that $\munderbar{\V{b}}_k$ determines $\Set{S}(\munderbar{\V{b}}_k) $. Furthermore, $(\munderbar{\V{b}}_k, m_k)$ jointly imply $m_k^{\fa} = m_k -\vert \Set{S}(\munderbar{\V{b}}_k) \vert $ whenever $m_k \geq \vert \Set{S}(\munderbar{\V{b}}_k) \vert$. Thus, the joint probability of $\munderbar{\RV{b}}_k$ and $\rv{m}_k $ becomes
\begin{IEEEeqnarray}{rll}
     p(&  \munderbar{\V{b}}_k, m_k \vert \munderbar{\V{x}}_k) \span \nonumber \\
    &=& p(\munderbar{\V{b}}_k, \Set{S}(\munderbar{\V{b}}_k), m_k \vert \munderbar{\V{x}}_k) \nonumber \\
    &=& p(\munderbar{\V{b}}_k \vert \Set{S}(\munderbar{\V{b}}_k),\munderbar{\V{x}}_k,  m_k) \, p(m_k \vert \Set{S}(\munderbar{\V{b}}_k), \munderbar{\V{x}}_k) \, p(\Set{S}(\munderbar{\V{b}}_k) \vert \munderbar{\V{x}}_k)  \nonumber \\
    &= &p(\munderbar{\V{b}}_k \vert \Set{S}(\munderbar{\V{b}}_k), m_k) p(m_k^{\fa}) p(\Set{S}(\munderbar{\V{b}}_k) \vert \munderbar{\V{x}}_k)  \nonumber \\
     &=& \Vuk{\Set{O}}(\munderbar{\V{x}}_k) e^{-\lfa} \frac{(\lfa)^{m_k^{\fa}}}{ m_{k}!}\, \Bigg[  \prod_{\Set{A}\in \Set{S}(\munderbar{\V{b}}_k)} \Pdk{\Set{A} }(\munderbar{\V{x}}_k^{\Set{A}})\,  \ruk{ \Set{A} } (\munderbar{\V{x}}_k^{\Set{A}}) \Bigg]  \nonumber  \\
    & & \times  \!\!\!\!  \sum_{\Set{U} \in \mathfrak{B}(\Set{S}(\V{b}_k^{(0)}))}  \prod_{\Set{S}\in \Set{U}}^{} \big( 1- \Pdk{\Set{S}}(\munderbar{\V{x}}_k^{\Set{S}})\big) \,  \ruk{\Set{S}} (\munderbar{\V{x}}_k^{\Set{S}} )    \nn\,.\label{eq_priorB_proof}
\end{IEEEeqnarray} 
The third equality above uses: (i) the independence of $\munderbar{\RV{b}}_k$ and $\munderbar{\RV{x}}_k$ when conditioned on the set of detected groups $\Set{S}(\munderbar{\V{b}}_k)$ and the number of measurements $m_k$; and (ii) the independence of false alarms from the detection of objects. Subsequently, in the last equality, we plugged in the expressions \eqref{eq_prob_misI} and \eqref{eq_p_det_groups}. 
 
\bibliographystyle{IEEEtran}
\bibliography{IEEEabrv,temp_clean}

\end{document}

%% file: Acronym-def.tex
\usepackage{acronym}

\acrodef{MIT}{Massachusetts Institute of Technology}
\acrodef{IIT}{Indian Institute of Technology}
\acrodef{IDSS}{Institute for Data, Systems, and Society }
\acrodef{LIDS}{Laboratory for Information and Decisions Systems}
\acrodef{WINSLAB}[WINSLab]{Wireless Information and Network Sciences Laboratory}

\acrodef{uwb}[UWB]{ultra-wideband}
\acrodef{ble}[BLE]{Bluetooth Low Energy}
\newacro{dfe}[DFE]{decision feedback equalizer}
\newacro{lms}[LMS]{least means square}
\newacro{sdn}[SDN]{software defined network}
\newacro{hca}[HCA]{heterogeneous computing architecture}
\acrodef{gps}[GPS]{Global Positioning System}
\acrodef{imu}[IMU]{inertial measurement unit}
\acrodef{csi}[CSI]{channel state information}
\acrodef{snr}[SNR]{signal-noise-ratio}
\acrodef{los}[LOS]{line-of-sight}
\acrodef{nlos}[NLOS]{non-line-of-sight}
\newacro{toa}[TOA]{time-of-arrival}
\newacro{tdoa}[TDOA]{time-difference-of-arrival}
\newacro{aoa}[AOA]{angle-of-arrival}
\newacro{rss}[RSS]{received signal strength}

\newacro{pocs}[POCS]{projection onto convex sets}
\newacro{eed}[EED]{Engineering Every Day}
\newacro{stem}[STEM]{science, technology, engineering, and mathematics}
\newacro{LOCC}[LOCC]{local operations and classical communication}
\newacro{ebit}[ebit]{entanglement bit}
\newacro{NISQ}[NISQ]{Noisy Intermediate-Scale Quantum}
\newacro{QSD}[QSD]{Quantum State Discrimination}
\newacro{amgm}[AMGM]{arithmetic mean geometric mean}
\newacro{wamgm}[WAMGM]{weighted arithmetic mean geometric mean}

\newacro{hdp}[HDP]{hierarchical Dirichlet process}
\newacro{CDF}[CDF]{Cumulative Distribution Function}
\newacro{crb}[CRB]{Cram\'{e}r-Rao Bound}
\newacro{speb}[SPEB]{squared position error bound}
\newacro{fim}[FIM]{Fisher information matrix}
\newacro{kld}[KLD]{Kullback--Leibler divergence}
\newacro{klda}[KLDA]{Kullback--Leibler divergence average}
\newacro{gaf}[GAF]{geometric--average fusion}
\newacro{pgaf}[PGAF]{particle geometric--average fusion}
\newacro{waf}[WAF]{weighted--average fusion}
\newacro{aaf}[AAF]{arithmetic--average fusion}
\newacro{ci}[CI]{covaraince intersection}
\newacro{dct}[DCT]{dominated convergence theorem}
\newacro{gdct}[GDCT]{generalized dominated convergence theorem}
\newacro{gmpgaf}[GM-PGAF]{Gaussian mixture~\ac{pgaf}}
\newacro{upgaf}[U-PGAF]{Uncorrected~\ac{pgaf}}
\newacro{bp}[BP]{Belief Propagation}
\newacro{gbp}[GBP]{Gaussian Belief Propagation}
\newacro{lbp}[LBP]{Loopy Belief Propagation}
\newacro{glbp}[GLBP]{Gaussian Loopy Belief Propagation}
\newacro{mc}[MC]{Monte Carlo}
\newacro{is}[IS]{importance sampling}
\newacro{pf}[PF]{particle filter}
\newacro{mis}[MIS]{multiple importance sampling}
\newacro{smc}[SMC]{sequential Monte Carlo}
\newacro{tvd}[TVD]{Total Variation Distance}
\newacro{atvd}[ATVD]{Average TVD}
\newacro{rl}[RL]{reinforcement learning}
\newacro{map}[MAP]{multi-agent planning}
\newacro{mdp}[MDP]{Markov decision process}
\newacro{pomdp}[POMDP]{partially observed Markov decision process}
\newacro{rpr}[RPR]{regionalized policy representation}
\newacro{irpr}[iRPR]{infinite regionalized policy representation}
\newacro{td}[TD]{temporal difference}
\newacro{rlmc}[RL-MC]{reinforcement learning Monte Carlo}
\newacro{dpomdp}[Dec-POMDP]{cecentralized-partially observed Markov decision process}
\newacro{fsc}[FSC]{finite state controller}

\newacro{ann}[ANN]{artificial neural network}
\newacro{dnn}[DNN]{deep neural network}
\newacro{em}[EM]{expectation maximization}

\newacro{mot}[MOT]{multi-object tracking}
\newacro{lmb}[LMB]{labeled multi-Bernoulli}
\newacro{mmlmb}[MM-LMB]{merged-measurement LMB}
\newacro{gmmlmb}[G-MM-LMB]{Gaussian MM-LMB}
\newacro{lrfs}[LRFS]{labeled random finite set}
\newacro{Lrfs}[LRFS]{Labeled random finite set}
\newacro{mou}[MOU]{measurement origin uncertainty}
\newacro{glmb}[GLMB]{generalized labeled multi-Bernoulli}
\newacro{rfs}[RFS]{random finite set}
\newacro{ospa}[OSPA]{optimum subpattern assignment}
\newacro{oot}[OoT]{ocean-of-things}
\newacro{Oot}[OoT]{Ocean-of-things}
\newacro{mot}[MOT]{multi-object tracking}
\newacro{ss}[SS]{sensor selection}
\newacro{ce}[CE]{cross-entropy}
\newacro{cess}[CE-SS]{cross-entropy SS}
\newacro{ootcess}[OoT-CESS]{OoT-CESS}
\newacro{da}[DA]{data-association}
\newacro{roi}[RoI]{region-of-interest}
\newacro{cs}[CS]{Cauchy–Schwarz}
\newacro{jpdaf}[JPDAF]{Joint Probabilistic Data Association Filter}
\newacro{jpda}[JPDA]{Joint Probabilistic Data Association}
\newacro{jpdam}[JPDAM]{Joint Probabilistic Data Association with Merged Measurements}
\newacro{mht}[MHT]{Multiple Hypothesis Tracking}
\newacro{imm}[IMM]{Interacting Multiple Models}
\newacro{utt}[UTT]{Unresolved Two-Target}
\newacro{mmse}[MMSE]{Minimum Squared Error}

\acrodef{uav}[UAV]{unmanned aerial vehicle}
\acrodef{nln}[NLN]{network localization and naviagation}
\newacro{ais}[AIS]{automatic identification system}
\acrodef{fy}[FY]{fiscal year}
\acrodef{iot}[IoT]{Internet of Things}
\newacro{aws}[AWS]{Amazon Web Services}
\newacro{alu}[ALU]{arithmetic logic unit}
\newacro{api}[API]{application programming interface}
\newacro{asic}[ASIC]{application-specific integrated circuit}